\documentclass{article}
\usepackage{bm,latexsym,amsmath,amsfonts,amssymb,fancyhdr,color,graphicx,slashed,cite}
\usepackage[a4paper,bottom=2.5cm,top=3.cm,head=5mm,width=17cm,dvipdfm]{geometry}
\usepackage[usenames,dvipsnames,svgnames,table]{xcolor}
\usepackage[colorlinks=true,
            linkcolor=blue,
            urlcolor=blue,
            citecolor=green,
            bookmarks=true,
            bookmarksnumbered=true,
            breaklinks=true,
            pdfpagemode=Fullscreen,
            pdfstartview=FitBH]{hyperref}

\usepackage[dotinlabels]{titletoc}
 \usepackage[normalem]{ulem}
\usepackage{float}
\usepackage{extarrows}
\usepackage{xcolor}
\usepackage{cancel}
\usepackage[bbgreekl]{mathbbol}
\usepackage{bbm}
\usepackage{siunitx}
\usepackage{enumitem}
\usepackage{tikz,pgfplots}
\usetikzlibrary{positioning}
\usepackage[title]{appendix}

\usepackage{multirow}
\usepackage{feynmf}
\numberwithin{equation}{section}
\allowdisplaybreaks[4]

\newcommand{\meqn}[1]{\hypersetup{linkcolor=blue}Eq.\,(\ref{#1})\hypersetup{linkcolor=blue}}
\newcommand{\mfig}[1]{{\hypersetup{linkcolor=violet}Fig.\,\ref{#1}\hypersetup{linkcolor=blue}}}
\newcommand{\mtab}[1]{{\hypersetup{linkcolor=purple}Tab.\,\ref{#1}\hypersetup{linkcolor=blue}}}

\newcommand{\td}{\mbox{d}}
\newcommand{\T}{\texttt{T}}
\newcommand{\C}{\texttt{C}}
\newcommand{\hc}{{\rm h.c.}}
\newcommand{\calL}{ {\cal L} }
\newcommand{\calO}{ {\cal O} }

\newcommand{\EoM}{\texttt{EoM}}
\newcommand{\IBP}{\texttt{IBP}}

\newcommand{\str}{{\rm STr}}

\usepackage[titles]{tocloft}

\begin{document}

\fontsize{10pt}{12pt}\selectfont

\title{\textbf{One-loop Matching of Scotogenic Model onto Standard Model Effective Field Theory up to Dimension 7} }

\author{
{Yi Liao~$^{a,b}$}\footnote{liaoy@scnu.edu.cn}\, and
{Xiao-Dong Ma~$^{c}$}\footnote{maxid@sjtu.edu.cn}
 \\[3mm]
{$^a$~Guangdong Provincial Key Laboratory of Nuclear Science, Institute of Quantum Matter,}\\
{South China Normal University, Guangzhou 510006, China}\\[1mm]
{$^b$~Guangdong-Hong Kong Joint Laboratory of Quantum Matter,}\\
{Southern Nuclear Science Computing Center,}\\
{South China Normal University, Guangzhou 510006, China}\\[1mm]
{$^c$~Tsung-Dao Lee Institute 
\& School of Physics and Astronomy
,} \\[1mm]
{Shanghai Jiao Tong University
, Shanghai 200240, China}
}
\date{}
\maketitle

\vspace{-0.75cm}

\begin{abstract}

The scotogenic neutrino seesaw model is a minimal extension of the standard model with three $\mathbb{Z}_2$-odd right-handed singlet fermions $N$ and one $\mathbb{Z}_2$-odd Higgs doublet $\eta$ that can accommodate the tiny neutrino mass and provide a dark matter candidate in a unified picture. Due to lack of experimental signatures for electroweak scale new physics, it is appealing to assume these new particles are well above the electroweak scale and take the effective field theory approach to study their effects on low energy observables. In this work we apply the recently developed functional matching formalism to the one-loop matching of the model onto the standard model effective field theory up to dimension seven for the case when all new states $N$ and $\eta$ are heavy to be integrated out. This is a realistic example which has no tree-level matching due to the $\mathbb{Z}_2$ symmetry. Using the matching results, we analyze their phenomenological implications for several physical processes, including the lepton number violating effect, the CDF $W$ mass excess, and the lepton flavor violating decays like $\mu\to e\gamma$ and $\mu\to 3e$.

\end{abstract}

\newpage
\hypersetup{linkcolor=black}
\tableofcontents
\hypersetup{linkcolor=red}

\section{Introduction}

The origin of neutrino mass and the nature of dark matter are two pieces of well-established evidence for physics beyond the standard model (SM). Among various beyond SM scenarios an attractive route is to correlate these two seemingly separate problems and solve them within the same theoretical framework. In this respect the scotogenic model~\cite{Ma:2006km} provides a nice example that can explain the origin of neutrino mass and dark matter (DM) in a single simple framework, with the introduction only of three $\mathbb{Z}_2$-odd right-handed singlet fermions $N$ and one $\mathbb{Z}_2$-odd scalar doublet $\eta$ on top of the SM content. The main merit of the model is that neutrinos gain radiative mass from interactions with particles in the dark sector. The lightest of the latter is stable due to the presumed exact $\mathbb{Z}_2$ symmetry, and thus potentially serves as a DM candidate. Since it was first proposed in 2006, the model has attracted a lot of attentions from various phenomenological aspects including the lepton flavor violating (LFV) golden modes $\mu \to e\gamma,~3e$ and $\mu$-$e$ conversion in nuclei~\cite{Kubo:2006yx, AristizabalSierra:2008cnr,Suematsu:2009ww,
Adulpravitchai:2009gi,Toma:2013zsa,Vicente:2014wga}, the scalar DM scenario~\cite{Huang:2018vcr,Avila:2019hhv,deBoer:2021pon,Avila:2021mwg},~
the fermionic WIMP and FIMP DM cases~\cite{Schmidt:2012yg, Ibarra:2016dlb,Hessler:2016kwm, Lindner:2016kqk,Baumholzer:2018sfb},
the running effect of neutrino masses \cite{Merle:2015ica} and the neutrino mass matrix textures~\cite{Kitabayashi:2018bye}, low scale leptogenesis~\cite{Baumholzer:2018sfb,Hugle:2018qbw,Borah:2018rca},
the signals at hadron and lepton colliders~\cite{Hessler:2016kwm,Baumholzer:2019twf, Liu:2022byu} and in
gravitational waves~\cite{Borah:2020wut}, and the LFV $Z$ and Higgs boson decays~\cite{Hundi:2022iva}, etc.

Nevertheless, there is no definite signature so far for the predicted particles on the observational side. It then looks natural to assume that the new particles lie well above the SM electroweak scale to be inaccessible to current high energy colliders. This motivates us to study their indirect effects on low energy observations by working with an effective field theory (EFT) in which they are integrated out. The standard model effective field theory (SMEFT) is tailored exactly for this purpose. It is a low energy effective field theory for SM particles below some new physics scale. In this framework the SM interactions appear as the leading term in a systematic series expansion, and the effects from new physics at a high scale are incorporated as suppressed high-dimensional operators and modifications to the SM parameters. The most appealing feature of SMEFT is perhaps its universality. It contains exclusively the SM fields which are governed by the SM gauge symmetries but is otherwise not constrained. Different high scale new physics will be reflected in Wilson coefficients (WCs) and their interrelations.
In the past years, the bases of complete and independent high dimensional operators have been built for the SMEFT up to dimension nine (dim-9)~\cite{Weinberg:1979sa,Grzadkowski:2010es,Lehman:2014jma, Henning:2015alf,Liao:2016hru,Li:2020gnx,Murphy:2020rsh,Li:2020xlh,Liao:2020jmn}. To apply the SMEFT to low energy phenomenology, one evolves it to the electroweak scale with the help of SMEFT renormalization group equations (RGE), matches it with the low energy effective field theory (LEFT) at the electroweak scale, and further evolves the latter to the experimental scale via the LEFT RGEs where one finally calculates the physical observables. In this way, low energy experimental data can be employed complementarily to constrain physics in the ultraviolet (UV).

An important task in this approach is the matching of a UV theory onto the SMEFT at the UV scale. Assuming the UV theory is perturbative, the matching is a double expansion, one in the number of loops and the other in the inverse power of the heavy scale. The tree-level matching can be easily done by solving the classical equations of motion (EoM) for the heavy fields followed by a low energy expansion to the desired order. However, in some cases interesting phenomenology (like flavor changing neutral currents) arises as a loop effect or precision data demands an improved theoretical analysis, so that one-loop matching becomes more and more relevant. Confronted with this, the recently developed functional matching via the effective action is a tailor-made method to achieve this goal~\cite{Henning:2014wua,Drozd:2015rsp,Henning:2016lyp,Ellis:2016enq,
Fuentes-Martin:2016uol,Zhang:2016pja,Ellis:2017jns,Kramer:2019fwz,
Cohen:2019btp,Cohen:2020fcu,Dittmaier:2021fls}. Unlike the diagrammatic approach, the matching is done by calculating some functional supertraces without computing Feynman diagrams one by one for a designed set of amplitudes. Some (semi-)automatic tools have been developed to facilitate this job for the tree-level matching~\cite{Criado:2017khh,
DasBakshi:2018vni,Carmona:2021xtq} and one-loop matching~\cite{Cohen:2020fcu,Cohen:2020qvb,Fuentes-Martin:2020udw,
Carmona:2021xtq}. One-loop matching has recently been practiced for several UV models, such as the three tree-level seesaws~\cite{Zhang:2021jdf,Coy:2021hyr,Ohlsson:2022hfl, Du:2022vso,Li:2022ipc}, the Zee model~\cite{Coy:2021hyr}, leptoquark models~\cite{Gherardi:2020det,Coy:2021hyr,Dedes:2021abc}, and others~\cite{Haisch:2020ahr,Jiang:2018pbd,Chala:2020vqp,Brivio:2021alv,Cepedello:2022pyx,Zhang:2022osj,Anisha:2021hgc}.

Considering rich phenomenology of the scotogenic model and null experimental searches for new heavy states, it is plausible to take the above EFT approach to investigate its low energy effects by treating both $N$ and $\eta$ as heavy states. It turns out that the existence of $\mathbb{Z}_2$ symmetry implies no tree-level matching to the SMEFT. Then we investigate its one-loop matching up to dim-7 operators. The direct result of functional evaluation is organized in the so-called Green basis with a handful of operators whose origin can be relatively easily tracked from the diagrammatic picture. To recast the result in terms of the standard dim-5 Weinberg operator~\cite{Weinberg:1979sa}, the dim-6 Warsaw basis~\cite{Grzadkowski:2010es} and the dim-7 basis~\cite{Liao:2016hru} (further improved in~\cite{Liao:2020roy}), some manipulations have to be made of the SM equations of motion (EoM), integration by parts (IBP) and the Fierz identities (FI).

The rest of the paper is organized as follows. In section \ref{sec:funcmatching} we review the basic ingredients of the functional matching in effective field theory. Section \ref{sec:scomodel} is a brief introduction of the scotogenic model as well as our notational convention. In section \ref{sec:1loopmatching} we consider the one-loop matching for the scotogenic model to the SMEFT, and show our main result in both Green and standard bases. The phenomenological analysis together with brief comparisons to the literature is included in section \ref{sec:pheno}. In section \ref{sec:conc}, we draw our conclusions. Supplementary materials presented in the appendices include the collection of the SMEFT operator bases up to dimension 7 in appendix~\ref{app:SMEFTbasis},  the calculation of the supertrace for four-lepton operators in appendix~\ref{sec:cal4fope}, and the reduction of operators from the Green basis to the standard basis in appendix~\ref{app:opered}.

\section{Basics of Functional Matching}
\label{sec:funcmatching}

For a UV field theory, whether fundamental or effective, with a hierarchical field spectrum, we collectively denote the heavy and light (scalar, fermion, or vector) fields as $\Phi$ and $\phi$, respectively, with a mass hierarchy $m_\Phi \gg m_\phi$. The low energy dynamics for light particles can be calculated either from the UV Lagrangian ${\cal L}_{\tt UV}(\Phi, \phi)$ consisting of both heavy and light fields, or from the EFT Lagrangian ${\cal L}_{\tt EFT}(\phi)$ consisting only of light fields. In matching calculation ${\cal L}_{\tt UV}(\Phi, \phi)$ is supposed to be known while ${\cal L}_{\tt EFT}(\phi)$ is searched for. To reproduce the low-energy physics of ${\cal L}_{\tt UV}(\Phi, \phi)$, ${\cal L}_{\tt EFT}(\phi)$ has to be carefully determined from ${\cal L}_{\tt UV}(\Phi, \phi)$ by integrating out $\Phi$ and performing a matching calculation. Conventionally, this matching is done by designing judiciously a complete set of amplitudes, computing them in both theories and equating them to determine the Wilson coefficients in ${\cal L}_{\tt EFT}(\phi)$ which depend on the parameters associated with $\Phi$. The main drawback of this diagrammatical matching is that one has to first determine the correct basis of operators at each dimension for the sought ${\cal L}_{\tt EFT}(\phi)$ and compute amplitudes twice. The procedure necessarily involves infrared physics of light particles which however eventually does not enter ${\cal L}_{\tt EFT}(\phi)$ itself, causing unnecessary complications.

A more elegant approach is the functional matching in the path integral formalism~\cite{Henning:2014wua, Drozd:2015rsp,Henning:2016lyp,Ellis:2016enq,Fuentes-Martin:2016uol,
Zhang:2016pja,Ellis:2017jns,Kramer:2019fwz,Cohen:2019btp,Cohen:2020fcu,
Dittmaier:2021fls}. The starting point is the identification of one-particle-irreducible (1PI)  generating functionals $\Gamma_{\tt EFT}[\phi] = \Gamma_{\tt UV}^{\tt L}[\phi]$ at the matching scale $m_\Phi$. Here $\Gamma_{\tt UV}^{\tt L}[\phi]$ is computed in the UV theory and irreducible only to the light field $\phi$ while $\Gamma_{\tt EFT}[\phi]$ is computed in the EFT. This identification is made in a double expansion, one in the inverse power expansion of $m_\Phi$ and the other in the number of loops. At the tree order, this is easy: one solves in the UV theory the classical EoM for $\Phi=\Phi_c[\phi]$ in terms of the light field $\phi$ which is a functional, and makes the inverse power expansion in $m_\Phi$ to turn it into an infinite series of local functions $\Phi=\Phi_c(\phi)$. Substituting it into ${\cal L}_{\tt UV}(\Phi,\phi)$ yields the answer: 
\begin{eqnarray}
{\cal L}_{\tt EFT}^{\tt tree}(\phi) 
= {\cal L}_{\tt UV}(\Phi_c(\phi), \phi).
\end{eqnarray}

To perform one-loop matching, let us start on the EFT side whose Lagrangian is 
\begin{eqnarray}
{\cal L}_{\tt EFT}(\phi) 
= {\cal L}_{\tt EFT}^{\rm tree}(\phi)
+  {\cal L}_{\tt EFT}^{\tt 1\mbox{-}loop}(\phi) + \cdots,
\end{eqnarray}
where the dots stand for higher-loop contributions and ${\cal L}_{\tt EFT}^{\tt 1\mbox{-}loop}(\phi)$ is what we are seeking. The above contributes to the one-loop 1PI generating functional in two manners, 
\begin{eqnarray}
\Gamma_{\tt EFT}^{\tt 1\mbox{-}loop}[\phi]  
=\int \td^d x ~{\cal L}_{\tt EFT}^{\rm 1\mbox{-}loop}(\phi)
+ \frac{i}{2}{\rm STr}[\ln(\pmb{\rm H}_{\tt EFT})],
\label{eq_1loopEFT}
\end{eqnarray}
where the second term arises from one-loop diagrams formed with interactions defined in ${\cal L}_{\tt EFT}^{\rm tree}(\phi)$. The supertrace (STr) includes a minus sign for fields that are quantized as a Grassmanian field. In the path integral formalism, it results from the Gaussian integral with the Hessian matrix being  
\begin{eqnarray}
\pmb{\rm H}_{\tt EFT}(x,y)
=\frac{\delta^2 S_{\tt EFT}^{\tt tree}[\phi]}{\delta\bar\phi(x)\delta\phi(y)},
\end{eqnarray}
where $S_{\tt a}^{\tt i}[\varphi]=\displaystyle\int \td^d z~{\cal L}_{\tt a}^{\tt i}(\varphi(z))$ denotes the action in $d$ dimensional spacetime at the ${\tt i}$ order in the ${\tt a}$ theory for the field $\varphi$. (A global factor of $\mu^{d-4}$ has been suppressed for brevity where $\mu$ is the usual renormalization scale.) To count correctly independent degrees of freedom, the fields have been  arranged in a self-conjugate form up to a rotation $R$, i.e., the conjugate pair of fields $\bar\varphi$ and $\varphi$ is related by $\bar\varphi=\varphi^\T R$ with $|{\rm det}(R)|=1$~\cite{Cohen:2020fcu,Cohen:2020qvb}. The supertrace term in \meqn{eq_1loopEFT} contains all infrared physics associated with the light field $\phi$. 

A similar 1-loop manipulation can be made for the generating functional $\Gamma_{\tt UV}^{\tt L}[\phi]$ in the UV theory. An important difference from the usual case is that it is a generating functional only for and irreducible only to the light field $\phi$ although the UV theory contains the heavy field $\Phi$ as well. Therefore when we make Legendre transform from the generating functional for connected Green's functions of $\phi$ to that for 1PI Green's functions $\phi$, we have to implement the classical EoM for the $\Phi$ field simply because no external source has been introduced to it. With this point in mind, we have, 
\begin{eqnarray}
\Gamma_{\tt UV}^{\tt  L,1\mbox{-}loop}[\phi]   
=  \frac{i}{2}{\rm STr} [\ln(\pmb{\rm H}_{\tt UV})],
\end{eqnarray}
where 
\begin{eqnarray}
\pmb{\rm H}_{\tt UV}(x,y)
=\left.\frac{\delta^2S_{\tt UV}[\varphi]}{\delta\bar\varphi(x)\delta\varphi(y)}
\right|_{\Phi=\Phi_c(\phi)}.
\end{eqnarray}
We notice that the Hessian matrix is defined for the whole field space $\varphi=(\phi,\Phi)$ and the substitution $\Phi=\Phi_c(\phi)$ is made only after the functional derivatives have been finished. 

The identification at the 1-loop order of $\Gamma_{\tt EFT}^{\tt 1\mbox{-}loop}[\phi]=\Gamma_{\tt UV}^{\tt  L,1\mbox{-}loop}[\phi]$ then implies 
\begin{eqnarray}
\int \td^d x ~{\cal L}_{\tt EFT}^{\rm 1\mbox{-}loop}(\phi)
=\frac{i}{2}{\rm STr} [\ln(\pmb{\rm H}_{\tt UV})]
-\frac{i}{2}{\rm STr}[\ln(\pmb{\rm H}_{\tt EFT})].
\label{eq_diff}
\end{eqnarray}
The recent development in functional matching is based on the following crucial realization~\cite{Henning:2016lyp,Cohen:2020fcu,Fuentes-Martin:2016uol, Zhang:2016pja}. When the loop integrals in the UV theory is calculated by integration by regions~\cite{Beneke:1997zp,Smirnov:2002pj}, 
\begin{eqnarray}
\frac{i}{2}{\rm STr} [\ln(\pmb{\rm H}_{\tt UV})]
=\left.\frac{i}{2}{\rm STr} [\ln(\pmb{\rm H}_{\tt UV})]\right|_{\tt hard}
+\left.\frac{i}{2}{\rm STr} [\ln(\pmb{\rm H}_{\tt UV})]\right|_{\tt soft},
\end{eqnarray}
the soft term exactly cancels out the EFT term in \meqn{eq_diff}. The calculation of one-loop matching then boils down to the calculation of the hard part in the UV term:
\begin{eqnarray}
\int \td^d x ~{\cal L}_{\tt EFT}^{\rm 1\mbox{-}loop}(\phi)
=\left.\frac{i}{2}{\rm STr} [\ln(\pmb{\rm H}_{\tt UV})]\right|_{\tt hard},
\end{eqnarray}
where the subscript hard means that the loop integrands are first Taylor-expanded for $q\sim m_\Phi \gg m_\phi, k$ where $q,~k$ stand for the loop and external momenta respectively, and then evaluated for the whole $q$ space in $d$ dimensions. 

The second functional derivative is split into two parts, $\pmb{\rm H}_{\tt UV}=\pmb{K} -\pmb{X}$, where $\pmb{K}$ contains only kinetic and mass terms and $\pmb{X}$ includes all remaining interactions. The matrix $\pmb{K}$ is in a block-diagonal form: 
\begin{equation}
K_i = \begin{cases}
P^2 - m_i^2 & \mbox{for spin-0 fields}\\
\slashed{P} - m_i & \mbox{for spin-$1/2$ fields}\\
-g^{\mu\nu}(P^2 - m_i^2)  & \mbox{for spin-1 fields in Feynman gauge}
\end{cases},
\end{equation}
where $P_\mu=i D_\mu$ with $D_\mu$ being the covariant derivative with respect to background gauge fields whose operators are under consideration. To proceed further, we make the Taylor expansion, 
\begin{equation}
\ln(\pmb{K} - \pmb{X}) 
=\ln(\pmb{K}) + \sum_{n=0}^\infty\frac{1}{n}(\pmb{K}^{-1}\pmb{X})^n. 
\end{equation}
Since $\pmb{X}$ contributes at least a mass dimension 1 ($3/2$) to operators in search when it involves a bosonic (fermionic) field and $\pmb{K}^{-1}$ contributes a nonnegative mass dimension upon finishing loop integrals, the expansion actually terminates for the sought operators with a given mass dimension. The evaluation of supertraces is then classified into a log-type and a  power-type, 
\begin{equation}
\int \td^d x {\cal L}_{\tt EFT}^{\tt 1\mbox{-}loop}[\phi_c] 
=\frac{i}{2}{\rm STr}\left[\ln ( \pmb{K} )\right]\Big|_{\tt hard}
-\frac{i}{2}\sum_{n=1}^{\infty} \frac{1}{n}{\rm STr}\left[\left(\pmb{K}^{-1}\pmb{X}\right)^n \right]\Big|_{\tt hard}.
\end{equation}
The log-type supertrace depends only on the representation in a gauge group and is thus universal. The evaluation of the supertraces is done by the technique of covariant derivative expansion (CDE)~\cite{Gaillard:1985uh,Chan:1986jq, Cheyette:1987qz}, which automatically leads to gauge  invariant operators. These methods have been implemented into semi-automatic tools like {\tt STrEAM}~\cite{Cohen:2020fcu,Cohen:2020qvb} and {\tt SuperTracer}~\cite{Fuentes-Martin:2020udw}. In this work, we will calculate both manually and with the help of {\tt SuperTracer} to achieve identical results.

\section{Review of Scotogenic Model}
\label{sec:scomodel}

We start first with the convention for the SM part. We denote the SM left-handed lepton and quark doublet fields as $L(1,2,1/2)$ and $Q(3,2,1/6)$, the right-handed up-type quark, down-type quark, and charged lepton singlet fields as $u(3,1,2/3)$, $d(3,1,-1/3)$, and $e(1,1,-1)$, and the Higgs doublet as $H(1,2,1/2)$, respectively. Here the numbers in brackets are the corresponding representations under the SM gauge group $SU(3)_C\times SU(2)_L\times U(1)_Y$, and the flavor index of the fermions is suppressed for simplicity. Dropping the gauge-fixing related terms and topological terms, the SM Lagrangian is, 
\begin{eqnarray}
\label{sml}
\mathcal{L}_{\tt SM}&=&
- \frac{1}{4}G^A_{\mu\nu}G^{A\mu\nu}
- \frac{1}{4}W^I_{\mu\nu}W^{I\mu\nu}
- \frac{1}{4}B_{\mu\nu}B^{\mu\nu}
\nonumber
\\
&&+ |D_\mu H|^2  +\mu_H^2 (H^\dagger H) - \lambda_H(H^\dagger H)^2
\nonumber
\\
&&+\sum_{\Psi=Q, L, u, d, e}\bar{\Psi}i \slashed{D}\Psi
-\left(\bar{Q}Y_u u \tilde{H}+\bar{Q}Y_d d H + \bar{L}Y_e e H +\mbox{h.c.}\right),
\label{eq:LSM}
\end{eqnarray}
where $\mu_H^2$ is related to the vacuum expectation value (vev) $v$ of the Higgs field via the relation, $\mu_H^2= \lambda_H v^2$, with $\lambda_H$ being the Higgs self-coupling consant.
The superscripts $A$ and $I$ count the generators of the groups $SU(3)_C$ and $SU(2)_L$, respectively. $Y_u,~Y_d,~Y_e$ are the Yukawa couplings which are complex matrices in flavor space, and $\tilde H_i=\epsilon_{ij}H^*_j$. The covariant derivative is defined by
\begin{eqnarray}
D_\mu=\partial_\mu-ig_3 T^AG^A_\mu-ig_2T^IW^I_\mu-ig_1YB_\mu,
\end{eqnarray}
where $g_{1},g_{2}$ and $g_{3} $ are the corresponding coupling constants for $U(1)_Y, SU(2)_L$, and $SU(3)_C$, and $T^A,~T^I,~Y$ are the generator matrices appropriate for the fields to be acted on.

The scotogenic model proposed by E. Ma  \cite{Ma:2006km} is the extension of the SM with three generations of the right-handed singlet fermion $N(1,1,0)$ and a second scalar doublet $\eta(1,2,1/2)$. A discrete $\mathbb{Z}_2$ symmetry is imposed under which these new fields are odd and the SM fields even. The complete Lagrangian of the scotogenic model takes the form~\cite{Ma:2006km}, 
\begin{eqnarray}
\calL &=& \calL_{\tt SM} + \calL_{N, \eta},
\nonumber
\\
\calL_{N, \eta}&=&\bar N i \slashed{\partial} N
- \left(\frac{1}{2}\overline{N} m_N N^\C + \bar L \tilde \eta Y_\eta N
+\hc  \right) + |D_\mu \eta|^2  - V(H, \eta),
\nonumber
\\
V(H, \eta) &=&  m_\eta^2 \eta^\dagger \eta
+ \frac{1}{2}\lambda_2 (\eta^\dagger \eta)^2
+ \lambda_3 (H^\dagger H) (\eta^\dagger \eta)
\nonumber
\\
&&+ \lambda_4 (H^\dagger \eta) (\eta^\dagger H)
+ \frac{1}{2}\lambda_5 [ (H^\dagger \eta)^2 +  (\eta^\dagger H)^2].
\label{eq:LSS}
\end{eqnarray}
The charge conjugation field is defined as $N^\C\equiv C\overline{N}^\T$ with $(N^\C)^\C=N$, where the charge conjugation matrix $C$ satisfies the relations $C^\T=C^\dagger = -C$ and $C^2=-1$.
$m_N$ is the Majorana mass matrix and assumed without loss of generality to be diagonal with real positive elements $m_{N_i}$, and $Y_\eta$ is the new Yukawa coupling that will enter the generation of neutrino mass. While the couplings $\lambda_{2,3,4}$ are real by themselves, the coupling  $\lambda_5$ can be chosen real by a phase redefinition of $\eta$, which only modifies the global phase in the Yukawa coupling $Y_\eta$. Assuming $\mu_H^2 > 0$ and $m_\eta^2 > 0$, when $H^0$ develops a vev as in the SM, $\eta^0$ is guaranteed not to because of the exact $\mathbb{Z}_2$ symmetry. The mixing between $H$ and $\eta$ is also forbidden by the symmetry.

In this work we assume both $\eta$ and $N$ particles are much heavier than the electroweak scale and we match the scotogenic model to the SMEFT at the scale $m_{\eta,N}$. Thus we will not shift the vev from the Higgs field $H$ to maintain the complete SM gauge symmetry. In particular, the whole $\lambda_{3,4,5}$ terms which would correct the mass of or even lift the degeneracy of the $\eta$ components upon including the vev, will be treated as perturbative interactions. This is indeed in accord with the EFT approach. 
Because of $\mathbb{Z}_2$ symmetry, the terms in $\calL $ are of even powers in $N$ and $\eta$, so that their classical EoMs are of odd powers. This means that they cannot be solved in terms of pure SM fields. Therefore, no higher dimensional operators can appear from tree-level matching in this case. This is obvious diagrammatically: pure $\mathbb{Z}_2$-even external lines cannot be connected at tree level with pure $\mathbb{Z}_2$-odd internal lines. The non-trivial matching then starts to appear at the one-loop level. For the other two possible cases in which one of $N$ and $\eta$ is treated as light and the other as heavy, tree-level matching indeed exists. For the case with a heavy $\eta$ and light $N$s, the model will match onto the sterile neutrino extended SMEFT ($\nu$SMEFT)~\cite{Liao:2016qyd}, with the lightest $N$ being DM. For the opposite case, it will match onto the DM EFT with a $\mathbb{Z}_2$-odd scalar doublet~\cite{Criado:2021trs}, of which the lightest neutral scalar could act as DM. These latter two cases have very different phenomenology and we defer the study of these possibilities to a future publication. 

\section{One-loop Matching onto the SMEFT}
\label{sec:1loopmatching}

To perform the one-loop matching using the functional method, we follow the notations in refs~\cite{Cohen:2020fcu,Cohen:2020qvb,Fuentes-Martin:2020udw} and introduce the pairs of fields that are self-conjugate up to a rotation matrix:
\begin{eqnarray}
&&\varphi_N \equiv N + N^\C, \quad
\varphi_\eta \equiv
\begin{pmatrix}
\eta \\ \eta^*
\end{pmatrix}.
\label{eq:varphi1}
\\
&&\bar \varphi_N = \varphi_N^\T C, \quad
\bar \varphi_\eta =  \varphi_\eta^\T
\begin{pmatrix}
0 & 1 \\
1 & 0
\end{pmatrix}=(\eta^\dagger,\eta^T).
\label{eq:varphi2}
\end{eqnarray}
This facilitates their path integral formulation as they appear like real variables. Let us first determine the interaction kernels $\pmb{X}_{ij}$ entering the power-type supertraces. 
The $\mathbb{Z}_2$ symmetry and that both $\eta$ and $N$s are heavy have the consequences: only the $\eta,N$ entries of $\pmb{X}$ and only the $\eta,N$-independent terms in those entries contribute to the matching. The former is because $\eta,N$ are odd and SM fields even under $\mathbb{Z}_2$, and the latter is due to absence of relevant terms from EoMs of $\eta,N$. Thus we only have the following three $\pmb{X}$s: 
\begin{subequations}
\label{eq:newX}
\begin{eqnarray}
\pmb{X}_{N\eta}^{[3/2]} & = &
\begin{pmatrix}
Y_\eta^\dagger  P_L (\epsilon^\T  L) & Y_\eta^\T  P_R (\epsilon^\T L^{\C})
\end{pmatrix},
\\%
\pmb{X}_{\eta N}^{[3/2]} & = &
\begin{pmatrix}
 (\bar L\epsilon) P_R  Y_\eta   \\
(\overline{L^\C}\epsilon) P_L  Y_\eta^*
\end{pmatrix},
\\%
\pmb{X}_{\eta\eta}^{[2],ij}  &= &
\begin{pmatrix}
\lambda_3 (H^\dagger H) \delta^{ij} + \lambda_4 H^i H^{j*}
&  \lambda_5 H^i H^j
\\
\lambda_5 H^{i*} H^{j*}
& \lambda_3 (H^\dagger H) \delta^{ij}+ \lambda_4 H^{i*} H^{j}
\end{pmatrix}
+\cdots,
\end{eqnarray}
\end{subequations}
where the superscript number in square brackets indicates the minimal canonical dimension of  $\pmb{X}$ and the dots stand for irrelevant $\eta$-dependent terms thus stated above.

\subsection{Matching result in a Green basis}

With the kernels in \meqn{eq:newX} at hand, we form the supertraces that induce the SMEFT operators up to dim 7: 
\begin{eqnarray}
\int d^d x \calL_{\tt EFT}^{\rm 1L,\,dim\leq 7} & =&
\left\{
{i \over 2} \str \log\pmb{K}
- {i \over 2}\str \left(\pmb{K}_\eta^{-1} \pmb{X}_{\eta\eta}^{[2]}\right)
- {1\over 2} {i \over 2}\str \left( \pmb{K}_\eta^{-1} \pmb{X}_{\eta\eta}^{[2]}\right)^2
-{1\over 3} {i \over 2}\str \left( \pmb{K}_\eta^{-1} \pmb{X}_{\eta\eta}^{[2]}\right)^3
\right.
\nonumber
\\
& - & {i \over 2}\str \left( \pmb{K}_\eta^{-1} \pmb{X}_{\eta N}^{[3/2]} \pmb{K}_N^{-1} \pmb{X}_{N\eta}^{[3/2]} \right)
-  {1\over 2}{i \over 2}\str \left(\pmb{K}_\eta^{-1} \pmb{X}_{\eta N}^{[3/2]}\pmb{K}_N^{-1} \pmb{X}_{N\eta}^{[3/2]} \right)^2
\nonumber
\\
& - &
 {i \over 2}\str \left(\pmb{K}_\eta^{-1} \pmb{X}_{\eta\eta}^{[2]}\pmb{K}_\eta^{-1} \pmb{X}_{\eta N}^{[3/2]}\pmb{K}_N^{-1} \pmb{X}_{N\eta}^{[3/2]} \right) 
\nonumber
\\
& - &\left.
{i \over 2}\str \left[ \left(\pmb{K}_\eta^{-1} \pmb{X}_{\eta\eta}^{[2]} \right)^2 \pmb{K}_\eta^{-1} \pmb{X}_{\eta N}^{[3/2]}\pmb{K}_N^{-1} \pmb{X}_{N\eta}^{[3/2]}\right]
 \right\}\Big|_{\tt hard}.
\label{eq:allsuper}
\end{eqnarray}
Note that symmetry factors have been included. 
An inspection of $\pmb{X}$ shows that the first six terms only contribute to the lepton-number-conserving (LNC) operators with an even mass dimension, the last but one term contributes to both LNC and lepton-number-violating (LNV) operators, and the last one only yields LNV dim-7 operators. 

\begin{table}
\centering
\resizebox{\linewidth}{!}{
\renewcommand{\arraystretch}{1.23}
{\footnotesize
\begin{tabular}{|c|c|c|}
\hline
dim &  Operator  &  WCs [$1/(16\pi^2)$] \\
\hline\hline%
     2
&  \cellcolor{red!15}$ H^\dagger H$
&  $(1+ L_\eta)(2\lambda_3 + \lambda_4)m_\eta^2$ \\
\hline\hline%
    \multirow{4}*{4}
&  \cellcolor{red!15}$B_{\mu\nu}B^{\mu\nu} \,(\blacksquare)$
&  $ - { L_\eta \over 24}g_1^2$      \\
 \cline{2-3}%
&   \cellcolor{red!15}$ W^{I\mu\nu}W^I_{\mu\nu}\,(\blacksquare)$
&   $ - {L_\eta \over 24}g_2^2$      \\
     \cline{2-3}%
 &  \cellcolor{red!15}$\overline{L_p}i \slashed{D} L_r$
 &  $\left[ Y_\eta { (1 + 2 L_\eta) m_\eta^4-4 (1+L_\eta)m_N^2 m_\eta^2 +(3 +2 L_N)m_N^4  \over 4(m_\eta^2 - m_N^2)^2} Y_\eta^\dagger \right]_{pr}$  \\
     \cline{2-3}%
 &  \cellcolor{red!15}$(H^\dagger H)^2$
 &  ${L_\eta\over 2}( 2\lambda_3^2+2 \lambda_3\lambda_4 +\lambda_4^2 +\lambda_5^2 )$  \\
 \hline\hline
   5
& $\epsilon_{im} \epsilon_{in} (\overline{L^{i\C}_p} L_r^j) H^m H^n +\hc $
& $ - \lambda_5   \left[Y_\eta^* m_N { m_\eta^2  - (1  - L_\eta + L_N )m_N^2
      \over 2 ( m_\eta^2 - m_N^2 )^2}Y_\eta^\dagger  \right]_{pr} $  \\
 \hline\hline
     \multirow{21}*{6}
 &  \cellcolor{blue!15}$\partial_\mu B^{\mu\nu} \partial^\rho B_{\rho\nu}\,(\blacksquare)$
 &  $-{g_1^2 \over 120 m_\eta^2 } $   \\
    \cline{2-3}%
&  \cellcolor{blue!15}$D_\mu W^{I\mu\nu} D^\rho W^I_{\rho\nu}\,(\blacksquare) ({\color{cyan}\divideontimes})$
&  $-{g_2^2 \over 120 m_\eta^2 }$  \\
     \cline{2-3}%
&  $\epsilon^{IJK} W_\mu^{I\nu}W_\nu^{J\rho} W_{\rho}^{K \mu}\,(\blacksquare)$
&  ${g_2^3\over 360 m_\eta^2 }$ \\
     \cline{2-3}%
&  $(H^\dagger H)^3$
&  $- { 2\lambda_3^3 +\lambda_4^3+ 3(\lambda_3+\lambda_4)(\lambda_3\lambda_4 +\lambda_5^2)
       \over 6 m_\eta^2}$ \\
     \cline{2-3}%
&  $ (H^\dagger H ) \partial^2 (H^\dagger H)$
&  $- { 2\lambda_3^2 + 2\lambda_3\lambda_4  -  \lambda_5^2 \over 12 m_\eta^2}$    \\
     \cline{2-3}%
&  $ |H^\dagger D_\mu H|^2$
&  $- {\lambda_4^2 - \lambda_5^2 \over 6 m_\eta^2}$ \\
     \cline{2-3}%
&  \cellcolor{blue!15}$ (H^\dagger H) (H^\dagger D^2 H) +\hc ({\color{cyan}\divideontimes}) $
&  $ - { \lambda_4^2  + \lambda_5^2 \over 12 m_\eta^2}$ \\
     \cline{2-3}%
&  $(H^\dagger H) B^{\mu\nu} B_{\mu\nu}$
&  ${g_1^2(2\lambda_3 + \lambda_4) \over 48 m_\eta^2}$ \\
     \cline{2-3}%
&  $ (H^\dagger H) W^{I \mu\nu} W^{I}_{\mu\nu}$
&  ${g_2^2(2\lambda_3 + \lambda_4) \over 48 m_\eta^2}$  \\
     \cline{2-3}%
&  $ (H^\dagger \sigma^I H) W^{I \mu\nu} B_{\mu\nu}$
&  ${g_1 g_2  \lambda_4 \over 24 m_\eta^2}$ \\
     \cline{2-3}%
&  \cellcolor{blue!15}$ (H^\dagger H)(\overline{L_p} i \overleftrightarrow{ \slashed{D}}  L_r)$
&  $ - (2\lambda_3 +\lambda_4) \left[ Y_\eta { m_\eta^4 -4 m_N^2 m_\eta^2 + (3 - 2L_\eta +2 L_N) m_N^4
       \over 8(m_\eta^2 - m_N^2 )^3} Y_\eta^\dagger \right]_{pr}$ \\
     \cline{2-3}%
&  \cellcolor{blue!15}$(H^\dagger \sigma^I H) (\overline{L_p} i \overleftrightarrow{ \slashed{D}}^I  L_r)$
&  $ \lambda_4 \left[ Y_\eta { m_\eta^4 -4 m_N^2 m_\eta^2 + (3 - 2L_\eta +2 L_N) m_N^4
      \over 8(m_\eta^2 - m_N^2)^3} Y_\eta^\dagger \right]_{pr}$ \\
     \cline{2-3}%
&  \cellcolor{blue!15}$\overline{L_p} i \overleftarrow{ \slashed{D}} \slashed{D} \slashed{D} L_r$
&  $\left[ Y_\eta {m_\eta^6 - 6m_N^2 m_\eta^4 + 3(1-2L_\eta + 2L_N)m_N^4 m_\eta^2 + 2m_N^6
      \over 6 (m_\eta^2 - m_N^2)^4} Y_\eta^\dagger \right]_{pr}$ \\
     \cline{2-3}%
&  \cellcolor{blue!15}$ B^{\mu\nu}  \overline{L_p} \sigma_{\mu\nu} i  \slashed{D} L_r +\hc$
&  $g_1\left[ Y_\eta {  m_\eta^6- 6m_N^2 m_\eta^4+ 3(1-2L_\eta + 2L_N)m_N^4 m_\eta^2  +2m_N^6
       \over 48(m_\eta^2 - m_N^2)^4} Y_\eta^\dagger \right]_{pr}$ \\
     \cline{2-3}%
&  \cellcolor{blue!15}$D_\nu B^{\mu\nu}  \overline{L_p} \gamma_\mu L_r$
&  $ - g_1 \left[ Y_\eta { 2 m_\eta^6 - 9m_N^2 m_\eta^4 +18 m_N^4 m_\eta^2 - (11- 6L_\eta + 6 L_N)m_N^6
       \over 72(m_\eta^2 - m_N^2)^4} Y_\eta^\dagger \right]_{pr}$ \\
     \cline{2-3}%
&  \cellcolor{blue!15} $W^{I\mu\nu}  \overline{L_p} \sigma^I \sigma_{\mu\nu} i  \slashed{D} L_r  +\hc$
&  $- g_2 \left[ Y_\eta {  m_\eta^6 - 6m_N^2 m_\eta^4 + 3(1-2L_\eta + 2L_N)m_N^4 m_\eta^2 +2m_N^6
      \over 48(m_\eta^2 - m_N^2)^4} Y_\eta^\dagger \right]_{pr}$ \\
     \cline{2-3}%
&  \cellcolor{blue!15}$D_\nu W^{I\mu\nu}  \overline{L_p}\sigma^I  \gamma_\mu L_r$
&  $ g_2 \left[ Y_\eta {2 m_\eta^6 - 9m_N^2 m_\eta^4 +18 m_N^4 m_\eta^2- (11- 6L_\eta + 6 L_N)m_N^6
      \over 72 (m_\eta^2 - m_N^2)^4} Y_\eta^\dagger \right]_{pr}$ \\
     \cline{2-3}%
&   \multirow{4}*{ $ (\overline{L_p}\gamma_\mu L_r) (\overline{L_s} \gamma^\mu L_t)$}
&   $\,\,\,- {1\over 4}\left\{ \left[ {(L_\eta - L_{N_v}) m_{N_v}^2 \over  (m_{N_v}^2 -m_{N_w}^2 )(m_\eta^2- m_{N_v}^2)^2} +v\leftrightarrow w\right] + {1\over  (m_\eta^2 - m_{N_v}^2) (m_\eta^2- m_{N_w}^2) } \right\}   $ \\
&
&  $\times   m_{N_v} m_{N_w} (Y_\eta)_{pw}  (Y_\eta^\T)_{ws} (Y_\eta^*)_{rv} (Y_\eta^\dagger)_{vt} $  \\
&
&  $- {1\over 8} \left\{ \left[ {(L_\eta - L_{N_v}) m_{N_v}^4  \over (m_{N_v}^2 -m_{N_w}^2 )(m_\eta^2- m_{N_v}^2)^2} + v\leftrightarrow w \right]
      + {m_\eta^2 \over  (m_\eta^2- m_{N_v}^2) (m_\eta^2 - m_{N_w}^2) }\right\}$ \\
 &
 &  $\times  (Y_\eta)_{pw}(Y_\eta^\dagger)_{wt}(Y_\eta)_{sv}  (Y_\eta^\dagger)_{vr} $  \\
\hline\hline
      \multirow{5}*{7}
&  $\epsilon_{im} \epsilon_{jn} (\overline{L^{i\C}_p} L_r^j) H^m H^n (H^\dagger H) + \hc$
&  $(\lambda_3 + \lambda_4)\lambda_5   \left[Y_\eta^*m_N {m_\eta^4+ 2(L_\eta - L_N) m_N^2 m_\eta^2
      -m_N^4  \over 2 m_\eta^2 (m_\eta^2 - m_N^2 )^3}Y_\eta^\dagger \right]_{pr}$ \\
     \cline{2-3}%
&  \cellcolor{blue!15}$\epsilon_{im} \epsilon_{jn} (\overline{L^{i\C}_p} D^2 L_r^j) H^m H^n + \hc $
&  $ \lambda_5  \left[Y_\eta^* m_N {m_\eta^4 + 4(1+ L_\eta - L_N ) m_N^2 m_\eta^2
     - (5 -2L_\eta + 2L_N ) m_N^4 \over 4 (m_\eta^2 - m_N^2 )^4}Y_\eta^\dagger  \right]_{pr}$  \\
     \cline{2-3}%
&  $\epsilon_{im} \epsilon_{jn}   (\overline{L^{i\C}_p} L_r^j) (D_\mu H^m D^\mu H^n) + \hc $
&  $ \lambda_5  \left[Y_\eta^* m_N{ m_\eta^6 - 6m_N^2 m_\eta^4 +3(1-2 L_\eta + 2L_N) m_N^4 m_\eta^2
       + 2m_N^6 \over 12m_\eta^2 (m_\eta^2 - m_N^2 )^4}Y_\eta^\dagger  \right]_{pr}$ \\
     \cline{2-3}%
&  \cellcolor{blue!15}$\epsilon_{im} \epsilon_{jn}   (\overline{L^{i\C}_p} L_r^j) H^m D^2 H^n +\hc $
&  $\lambda_5  \left[Y_\eta^* m_N { m_\eta^6 - 6m_N^2 m_\eta^4 +3(1-2 L_\eta + 2L_N) m_N^4 m_\eta^2
       + 2m_N^6 \over 12m_\eta^2 (m_\eta^2 - m_N^2 )^4} Y_\eta^\dagger  \right]_{pr}$ \\
\hline
\end{tabular} } }
\caption{The one-loop matching result for the scotogenic model in the heavy $N$ and $\eta$ case in a Green basis. The pink sector contributes to the threshold correction, while the blue sector will be reduced by using EoMs. See the text for the notations $\blacksquare$ and ${\color{cyan}\divideontimes}$.}
\label{tab:radssaw_Gbasis}
\end{table}

The supertraces are calculated in a gauge invariant way by the covariant derivative expansion method~\cite{Henning:2014wua,Henning:2016lyp}. It has been incorporated into partially automatic {\it Mathmatica} packages like {\tt STrEAM}~\cite{Cohen:2020qvb} and  {\tt SuperTracer}~\cite{Fuentes-Martin:2020udw}. For practical purposes, we have computed these supertraces both with the help of the  {\tt SuperTracer} package and manually as a crosscheck. In  \mtab{tab:radssaw_Gbasis}, we first provide the final matching result up to dim 7 in a minimal Green basis without implementing the field redefinitions or EoMs to go back into the standard basis~\cite{Grzadkowski:2010es,Liao:2020roy}. Such a basis are helpful in that one can relatively easily track the origin of the matched operators from the diagrammatic viewpoint. In addition, a handful of operators appear for 
the matching at each dimension. For the result in \mtab{tab:radssaw_Gbasis}, several comments are in order:
\begin{enumerate}[topsep=0ex]
\setlength\itemsep{0em} %
\item We have defined $L_\eta \equiv\epsilon^{-1}+\ln(4\pi\mu^2/m_\eta^2)-\gamma_{\rm E}$ and $L_N\equiv\epsilon^{-1}+\ln(4\pi\mu^2/m_N^2)-\gamma_{\rm E}$, which are related to the UV divergence in the modified minimal subtraction scheme in $d=4-2\epsilon$ dimensions; 
\item The log-type supertrace is only for the scalar $\eta$ due to its non-trivial $SU(2)_L\times U(1)_Y$ representation, and it generates the operators consisting exclusively of gauge fields that are marked by a $\blacksquare$;
\item The IBP relations and group $SU(2)_L$ and four-fermion Fierz identities are extensively used to achieve the operators in the table. For instance, for the dim-6 operators with pure Higgs fields and a pair of derivatives, we have used the following transformations to reach the operators in the table,
\begin{subequations}
\label{eq:scalaropered}
\begin{eqnarray}
 (H^\dagger H)   |D_\mu H|^2 & \overset{\IBP}{\Rightarrow}&
{1\over 2}(H^\dagger H ) \partial^2 (H^\dagger H)
-{1\over 2} [ (H^\dagger H) (H^\dagger D^2 H)+\hc],
\\%
 (H^\dagger D_\mu H)(H^\dagger D^\mu H)
&\overset{\IBP}{\Rightarrow}  &
 - {1\over 2}(H^\dagger H ) \partial^2 (H^\dagger H)
 - |H^\dagger D_\mu H|^2
\nonumber
\\
&& -{1\over 2} [ (H^\dagger H) (H^\dagger D^2 H) - \hc],  \quad\quad
\\%
(H^\dagger i \overleftrightarrow{D_\mu} H) (H^\dagger i \overleftrightarrow{D^\mu} H)
&\overset{\IBP}{\Rightarrow}&
(H^\dagger H ) \partial^2 (H^\dagger H)+4 |H^\dagger D_\mu H|^2,
\\%
(H^\dagger i \overleftrightarrow{D_\mu}^I H)(H^\dagger i \overleftrightarrow{D^\mu}^I H)
& \overset{\IBP}{\Rightarrow} & 3 (H^\dagger H) \partial^2 (H^\dagger H)
-2 [ (H^\dagger H) (H^\dagger D^2 H) +\hc].
\end{eqnarray}
\end{subequations}
\item A WC in the form, $[Y_\eta f(m_N) Y_\eta^\dagger]_{pr}$, should be understood as a matrix multiplication, $(Y_\eta)_{px} f(m_{N_x})(Y_\eta^\dagger)_{xr}$ with the dummy index $x$ of $N_x$ being summed over. In other words, the factor sandwiched between the two Yuwaka matrices is understood as a diagonal matrix with the $x$-th diagonal element evaluated at $m_{N_x}$. 
The same notation is used throughout the paper.
\item The four-lepton operators in the last row of dim-6 sector are generated by the 5th power-type supertrace in~\meqn{eq:allsuper}. Since the current version of {\tt SuperTracer} package cannot deal with the non-degenerate fermion case in a fully automatic way, we demonstrate our manual calculation in appendix~\ref{sec:cal4fope}. In the degenerate mass limit, we find our result is consistent with the output from using~{\tt SuperTracer}.
\item The dim-4 or less operators in the pink sector give the so-called threshold corrections to the SM, and they lead to the renormalization of the SM parameters. The dim-6 and dim-7 operators in blue are not yet in the standard basis~\cite{Grzadkowski:2010es,Liao:2020roy}
and will be carefully dealt with in the following subsection to obtain the final matching result in the standard basis.
\item An interesting feature for the matching result in \mtab{tab:radssaw_Gbasis} is that UV divergence only appears in the WCs of dim$\leq 4$ operators (characterized by the $L_\eta$) and cancels out for dim$\geq 5$ operators in a combination form, $L_\eta - L_N = \ln (m_N^2/m_\eta^2)$. This is as expected, since a correct implementation of renormalization should guarantee that all UV divergence in a renormalizable theory like the scotogenic model can always be absorbed into its parameters associated with dim$\leq 4$ operators. 
 
\end{enumerate}

\subsection{Matching result in the standard basis}

To translate the matching result in the Green basis in \mtab{tab:radssaw_Gbasis} into one in the standard dim-6~\cite{Grzadkowski:2010es} and dim-7 basis~\cite{Liao:2020roy}, we need to tackle both the threshold correction to the SM terms in \meqn{eq:LSM} and the reduction of those higher dimensional operators not in the standard basis by field or coupling redefinitions.
For this purpose one first renormalizes the SM terms, derives EoMs from the renormalized SM Lagrangian and applies them to the reduction of higher dimensional operators. 
From the pink sector in \mtab{tab:radssaw_Gbasis} the modified SM terms are, 
\begin{eqnarray}
\calL_{\rm SM}&\supset& 
- {1\over 4} (1 + \delta Z_B) B_{\mu\nu} B^{\mu\nu}
- {1\over 4}(1 + \delta Z_W) W_{\mu\nu}^I W^{I\mu\nu}
\nonumber
\\
&&+  \overline{L} (1+ \delta Z_L) i\slashed{D}L
+ \hat \mu_H^2 H^\dagger H
- \hat \lambda_H (H^\dagger H)^2,
\end{eqnarray}
where the field renormalization constants and modified parameters in the Higgs potential due to one-loop matching are given by 
\begin{subequations}
\label{eq:SMcorrection}
\begin{eqnarray}
\delta Z_B & = & {1\over 16\pi^2 } {L_\eta \over 6} g_1^2,
\\
\delta Z_W & = & {1\over 16\pi^2 } { L_\eta \over 6} g_2^2,
\\
(\delta Z_L)_{pr} & = & {1\over 16\pi^2 } \left[ Y_\eta {(3 +2 L_N)m_N^4 -4 (1+L_\eta)m_N^2 m_\eta^2 +(1 + 2 L_\eta) m_\eta^4 \over 4(m_N^2 - m_\eta^2)^2} Y_\eta^\dagger \right]_{pr},
\\
\hat \mu_H^2 &  =  & \mu_H^2 +  {1\over 16\pi^2 }  (1+ L_\eta)(2\lambda_3 + \lambda_4)m_\eta^2,
\\
\hat \lambda_H &=& \lambda_H
- {1\over 16\pi^2}{L_\eta\over 2}( 2\lambda_3^2+2 \lambda_3\lambda_4 +\lambda_4^2 +\lambda_5^2 ), 
\end{eqnarray}
\end{subequations}
with $\delta Z_L^\dagger =\delta Z_L$.
To bring kinetic terms into the canonical form, we make the following field and coupling  redefinitions,
\begin{subequations}
\label{eq:SMnor}
\begin{eqnarray}
&&B_\mu \to \left(1 -{1\over 2}\delta Z_B \right)B_\mu, \,
W^I_\mu \to \left(1 -{1\over 2}\delta Z_B\right) W^I_\mu,\,
L_p \to \left(1 -{1\over 2}\delta Z_L\right)_{pr} L_r,
\\
&&
\,\, g_1  \to \left(1 + {1\over 2}\delta Z_B \right) g_1, \,\quad
g_2 \to \left(1 + {1\over 2}\delta Z_B\right) g_2.
\end{eqnarray}
\end{subequations}
The gauge coupling redefinitions ensure the covariant derivative is unchanged under gauge field redefinitions. These redefinitions restore the SM Lagrangian in its canonical form but with modified lepton Yukawa coupling and Higgs potential parameters, 
\begin{eqnarray}
\label{sml}
\mathcal{L}_{\rm SM}^{\rm R}&=&
- \frac{1}{4}G^A_{\mu\nu}G^{A\mu\nu}
- \frac{1}{4}W^I_{\mu\nu}W^{I\mu\nu}
- \frac{1}{4}B_{\mu\nu}B^{\mu\nu}
\nonumber
\\
&&+ |D_\mu H|^2  +\hat \mu_H^2 (H^\dagger H) - \hat \lambda_H(H^\dagger H)^2
\nonumber
\\
&&+\sum_{\Psi}\bar{\Psi}i \slashed{D}\Psi
-\left[\bar{Q}Y_u u \tilde{H}+\bar{Q}Y_d d H + \bar{L}\hat Y_e e H +\mbox{h.c.}\right],
\end{eqnarray}
where the renormalized $\hat \mu_H^2$ and $\hat \lambda_H$ are given in \meqn{eq:SMcorrection} and the renormalized lepton Yuwaka coupling is
\begin{eqnarray}
(\hat Y_e)_{pr} = (Y_e)_{pr} - {1\over 2}(\delta Z_L)_{ps}(Y_e)_{sr}.
\label{eq:SMLYu}
\end{eqnarray}
The above redefinition or renormalization also modifies the one-loop generated higher dimensional operators, but the effect is a two-loop correction and can thus be neglected in our one-loop  matching.

We can now derive the EoMs from the renormalized SM Lagrangian $\mathcal{L}_{\rm SM}^{\rm R}$ which are the usual ones with the substitutions,  $(\mu^2_H,~\lambda_H,~Y_e)\to(\hat\mu^2_H,~\hat\lambda_H,~\hat Y_e)$. For the reduction of operators  in the blue sector in \mtab{tab:radssaw_Gbasis}, only the following ones are required, 
\begin{subequations}
\label{eq:SMEoM}
\begin{eqnarray}
\partial^\nu B_{\mu\nu} &=&
g_1
\left[
{1\over 6} \bar{Q} \gamma_\mu Q
+ {2\over 3} \bar{u} \gamma_\mu u
- {1\over 3} \bar{d} \gamma_\mu d
- {1\over 2} \bar{L} \gamma_\mu L
- \bar{e} \gamma_\mu e
+ {1\over 2}  H^\dagger i \overleftrightarrow{D_\mu} H
 \right],
 \\ %
D^\nu W^I_{\mu\nu} &=& { g_2 \over 2}
\left[  \overline{Q}\sigma^I\gamma_\mu Q
+  \overline{L}\sigma^I\gamma_\mu L
 + H^\dagger  i \overleftrightarrow{D_\mu}^I H
 \right],
\\%
D^2H &=& \hat \mu_H^2H-2  \hat \lambda_H (H^\dagger H)H-\epsilon^{\T}\bar{Q}  Y_u u-\bar{d}Y^\dagger_dQ-\bar{e}Y^\dagger_eL ,
\label{eq:Heom}
\\%
(D^2H)^\dagger &=&\hat \mu_H^2 H^\dagger-2 \hat \lambda_H (H^\dagger H)H^\dagger-\bar{u}  Y_u^\dagger Q\epsilon-\bar{Q}Y_d d-\bar{L}\hat Y_e e ,
\\%
 i\slashed{D}L & = &\hat Y_e e H,
\\
-i\bar{L}\overleftarrow{\slashed{D}} &=& H^\dagger\bar{e} Y_e^\dagger,
\\%
i\slashed{D}e&=&\hat  Y^\dagger_e {H}^\dagger  L,
\\
-i\bar{e}\overleftarrow{\slashed{D}}&= & \bar{L}\hat Y_eH.
\end{eqnarray}
\end{subequations}
Once again since the higher dimensional operators are generated at one-loop order, the hat in the three parameters can be dropped causing a negligible two-loop error in the result. 

\begin{table}
\centering
\resizebox{\linewidth}{!}{
\renewcommand{\arraystretch}{1.25}
{\footnotesize
\begin{tabular}{|c|c|c|}
\hline
Class &  Operator  &  WCs [$1/(16\pi^2 m_\eta^2)$] \\
\hline\hline%
$X^3$
& $\calO_{W}$
& $ {1 \over 360 } g_2^3$ \\
\hline%
$H^6$
& $\calO_{H}$
& $- {1\over 6} [2\lambda_3^3 +\lambda_4^3+ 3(\lambda_3+\lambda_4)(\lambda_3\lambda_4 +\lambda_5^2)]
       +{1 \over 60}[ 20 (\lambda_4^2 +\lambda_5^2)- g_2^4] \lambda_H  $  \\
\hline%
\multirow{2}*{$H^4D^2$}
& $\calO_{H\Box}$
&  $- { 1 \over 12 } (2\lambda_3^2 + 2\lambda_3\lambda_4  -  \lambda_5^2)
      - {1 \over 480} (g_1^4 +3g_2^4)
     $   \\
\cline{2-3}
& $\calO_{HD}$
& $- {1 \over 6}(\lambda_4^2 - \lambda_5^2)
    - {1\over 120} g_1^4
    $ \\
\hline%
\multirow{3}*{$ \psi^2H^3+\hc$}
&  $\calO_{eH}^{pr}$
&  $  {1\over 240 } [20(\lambda_4^2 +\lambda_5^2) -g_2^4 ] (Y_e)_{pr}
       - {1\over 4}\lambda_3 \left[ Y_\eta F_1(x) Y_\eta^\dagger  Y_e\right]_{pr}
       +  {1\over 12}\left[Y_e Y_e^\dagger Y_\eta F_2(x) Y_\eta^\dagger Y_e\right]_{pr}  $ \\
\cline{2-3}
& \cellcolor{blue!15}$\calO_{uH}^{pr}$
& $ {1 \over 240 }[20(\lambda_4^2 +\lambda_5^2) -g_2^4](Y_u)_{pr}  $ \\
\cline{2-3}
& \cellcolor{blue!15} $\calO_{dH}^{pr}$
& $ {1 \over 240 }[20(\lambda_4^2 +\lambda_5^2) -g_2^4] (Y_d)_{pr}  $ \\
\hline
\multirow{3}*{$X^2H^2$}
& $\calO_{HW}$
&  ${1\over 48} g_2^2(2\lambda_3 + \lambda_4) $  \\
\cline{2-3}
& $\calO_{HB}$
& ${1 \over 48} g_1^2(2\lambda_3 + \lambda_4)$ \\
\cline{2-3}
& $\calO_{HWB}$
& ${1 \over 24} g_1 g_2  \lambda_4 $ \\
\hline
\multirow{2}*{$ \psi^2XH+\hc$}
& \cellcolor{blue!15}$\calO_{eW}^{pr}$
&  $ - {1\over 48} g_2 \left[ Y_\eta F_2(x)Y_\eta^\dagger Y_e\right]_{pr}$ \\
\cline{2-3}
& \cellcolor{blue!15}$\calO_{eB}^{pr}$
& $ ~~ {1\over 48} g_1  \left[ Y_\eta F_2(x) Y_\eta^\dagger Y_e\right]_{pr}$ \\
\hline
\multirow{7}*{$\psi^2H^2D$}
&  \cellcolor{blue!15} $\calO_{Hl}^{(1),pr}$
&  $~~\, {1\over 240} g_1^4  \delta_{pr}  - {1\over 72}g_1^2 \left[ Y_\eta F_3(x) Y_\eta^\dagger \right]_{pr}$  \\
\cline{2-3}%
&  \cellcolor{blue!15}$\calO_{Hl}^{(3),pr}$
& $ - {1 \over 240} g_2^4 \delta_{pr} +  {1\over 72} g_2^2  \left[ Y_\eta F_3(x) Y_\eta^\dagger \right]_{pr}$ \\
\cline{2-3}%
& \cellcolor{blue!15} $\calO_{He}^{pr}$
& $~~~~~~ {1 \over 120 } g_1^4  \delta_{pr} + {1\over 12} \left[Y_e^\dagger Y_\eta F_2(x) Y_\eta^\dagger Y_e\right]_{pr}$ \\
\cline{2-3}%
& \cellcolor{blue!15} $\calO_{Hq}^{(1),pr}$
& $- {1 \over 720}g_1^4 \delta_{pr} $ \\
\cline{2-3}%
& \cellcolor{blue!15} $\calO_{Hq}^{(3),pr}$
&  $ - {1 \over 240}g_2^4 \delta_{pr} $ \\
\cline{2-3}%
& \cellcolor{blue!15} $\calO_{Hu}^{pr}$
&   $- {1 \over 180}g_1^4 \delta_{pr} $ \\
\cline{2-3}%
& \cellcolor{blue!15} $\calO_{Hd}^{pr}$
&  $~~ {1\over 360}g_1^4 \delta_{pr} $ \\
\hline
\multirow{7}*{$(\bar LL)(\bar LL)$}
&   \multirow{3}*{ $\calO_{ll}^{prst}$}
&   $F_{4,1}(x_v,x_w) (Y_\eta)_{pw}  (Y_\eta^\T)_{ws} (Y_\eta^*)_{rv} (Y_\eta^\dagger)_{vt}
 - F_{4,2}(x_v,x_w)  (Y_\eta)_{pw}(Y_\eta^\dagger)_{wt}(Y_\eta)_{sv}  (Y_\eta^\dagger)_{vr}$ \\
  &
&  $ +    {1\over 144} \left\{  (g_1^2 - g_2^2) \left[ Y_\eta F_3(x) Y_\eta^\dagger \right]_{pr} \delta_{st}
 + 2 g_2^2 \left[ Y_\eta F_3(x) Y_\eta^\dagger \right]_{pt} \delta_{rs}
+(pr) \leftrightarrow (st) \right\}$ \\
&
&      $   - {1\over 480}[ (g_1^4 - g_2^4)\delta_{pr} \delta_{st} + 2 g_2^4 \delta_{pt} \delta_{rs}] $ \\
\cline{2-3}%
& \cellcolor{blue!15}$\calO_{qq}^{(1),prst}$
& $ - {1 \over 4320} g_1^4\delta_{pr} \delta_{st} $ \\
\cline{2-3}%
& \cellcolor{blue!15}$\calO_{qq}^{(3),prst}$
& $\,\, - { 1\over 480}g_2^4 \delta_{pr} \delta_{st} $ \\
\cline{2-3}%
& \cellcolor{blue!15}$\calO_{lq}^{(1),prst}$
& $~~~~ {1 \over 720} g_1^4\delta_{pr} \delta_{st}
       - {1\over 216} g_1^2 \left[ Y_\eta F_3(x) Y_\eta^\dagger \right]_{pr} \delta_{st}$ \\
\cline{2-3}%
& \cellcolor{blue!15}$\calO_{lq}^{(3),prst}$
& $  -{1\over 240} g_2^4 \delta_{pr} \delta_{st}
        + {1\over 72} g_2^2 \left[ Y_\eta F_3(x) Y_\eta^\dagger \right]_{pr} \delta_{st}$ \\
\hline
\multirow{6}*{$(\bar RR)(\bar RR)$}
& \cellcolor{blue!15}$\calO_{ee}^{prst}$
&  $\,\,\, - { 1\over 120}g_1^4 \delta_{pr} \delta_{st} $ \\
\cline{2-3}%
& \cellcolor{blue!15}$\calO_{uu}^{prst}$
& $\,\,\, - { 1\over 270}g_1^4 \delta_{pr} \delta_{st} $ \\
\cline{2-3}%
& \cellcolor{blue!15}$\calO_{dd}^{prst}$
& $ - { 1\over 1080}g_1^4 \delta_{pr} \delta_{st} $ \\
\cline{2-3}%
& \cellcolor{blue!15}$\calO_{eu}^{prst}$
& $~~~~~  { 1\over 90}g_1^4 \delta_{pr} \delta_{st} $ \\
\cline{2-3}%
& \cellcolor{blue!15}$\calO_{ed}^{prst}$
& $\,\,\, - { 1\over 180}g_1^4 \delta_{pr} \delta_{st} $\\
\cline{2-3}%
& \cellcolor{blue!15}$\calO_{ud}^{(1),prst}$
& $~~~~  { 1\over 270}g_1^4 \delta_{pr} \delta_{st} $ \\
\hline%
\multirow{8}*{$(\bar LL)(\bar RR)$}
& \cellcolor{blue!15}$\calO_{le}^{prst}$
&  $\,\,\, -  {1\over 120 }g_1^4  \delta_{pr} \delta_{st}
      + {1\over 36}g_1^2 \left[ Y_\eta F_3(x) Y_\eta^\dagger \right]_{pr} \delta_{st}$ \\
\cline{2-3}%
&  \cellcolor{blue!15}$\calO_{lu}^{prst}$
& $\,\,\,\,\,\,\, {1 \over 180 } g_1^4  \delta_{pr} \delta_{st}
       - {1\over 54 }g_1^2  \left[ Y_\eta F_3(x) Y_\eta^\dagger \right]_{pr} \delta_{st}$ \\
\cline{2-3}%
&  \cellcolor{blue!15}$\calO_{ld}^{prst}$
& $\,\,\,\,- {1\over 360 }g_1^4  \delta_{pr} \delta_{st}
       +{1\over 108} g_1^2 \left[ Y_\eta F_3(x) Y_\eta^\dagger \right]_{pr} \delta_{st}$ \\
\cline{2-3}%
&  \cellcolor{blue!15}$\calO_{qe}^{prst}$
& $~~ {1 \over 360}g_1^4 \delta_{pr} \delta_{st}$ \\
\cline{2-3}%
&  \cellcolor{blue!15}$\calO_{qu}^{(1),prst}$
& $- {1\over 540}g_1^4  \delta_{pr} \delta_{st}$ \\
\cline{2-3}%
&  \cellcolor{blue!15}$\calO_{qd}^{(1),prst}$
& $~~ {1\over 1080}g_1^4  \delta_{pr} \delta_{st}$ \\
 \hline
\end{tabular}
}
}
\caption{The matching result of dim-6 operators in the standard basis~\cite{Grzadkowski:2010es}.}
\label{tab:radssaw_Wbasis}
\end{table}

\begin{table}
\centering
\resizebox{\linewidth}{!}{
\renewcommand{\arraystretch}{2}
{\footnotesize
\begin{tabular}{|c|c|c|}
\hline
Class &  Operator  &  WCs [$1/(16\pi^2)$] \\
\hline\hline%
$ \psi^2 H^2$
& $\calO_{LH,5}^{pr}$
&  $- {\lambda_5 \over 2m_\eta^2}\left\{  \left[Y_\eta^* m_N G_1(x)Y_\eta^\dagger  \right]_{pr}
      - {\mu_H^2 \over 6 m_\eta^2}  \left[Y_\eta^* m_N G_4(x) Y_\eta^\dagger  \right]_{pr} \right\}$ \\
\hline \hline%
$ \psi^2 H^4$
& $\calO_{LH}^{pr}$
& $ {\lambda_5 \over 2 m_\eta^4}\left\{  (\lambda_3 + \lambda_4) \left[Y_\eta^* m_N G_2(x) Y_\eta^\dagger \right]_{pr}
   -  {\lambda_H \over 3} \left[ Y_\eta^* m_N G_4(x) Y_\eta^\dagger \right]_{pr} \right\}$ \\
\hline
$ \psi^2 H^3D $
&\cellcolor{blue!15} $\calO_{LeHD}^{pr}$
& $\, { \lambda_5 \over 4 m_\eta^4}  \left[Y_\eta^*m_N G_3(x) Y_\eta^\dagger  Y_e \right]_{pr}$  \\
\hline
$ \psi^2 H^2 X $
&\cellcolor{blue!15} $\calO_{LHW}^{pr}$
&  $ - { \lambda_5 \over 16 m_\eta^4}  \left[Y_\eta^*m_N G_3 (x)Y_\eta^\dagger  \right]_{pr}$  \\
\hline
 $ \psi^2 H^2D^2 $
& $\calO_{LDH2}^{pr}$
& \,\,\, $  { \lambda_5 \over 12 m_\eta^4}  \left[Y_\eta^* m_N G_4(x) Y_\eta^\dagger   \right]_{pr}$ \\
\hline
\multirow{3}*{ $ \psi^4 H$ }
&\cellcolor{blue!15} $\calO_{\bar eLLLH}^{prst}$
& \,\, $ { \lambda_5 \over 12 m_\eta^4} (Y_e^\dagger)_{pr} \left[Y_\eta^* m_N G_4(x)Y_\eta^\dagger  \right]_{st}$ \\
\cline{2-3}%
&\cellcolor{blue!15} $\calO_{\bar dQLLH1}^{prst}$
& \,\, $ { \lambda_5 \over 12 m_\eta^4} (Y_d^\dagger)_{pr} \left[Y_\eta^* m_N G_4(x) Y_\eta^\dagger  \right]_{st}$ \\
\cline{2-3}%
&\cellcolor{blue!15} $\calO_{\bar QuLLH}^{prst}$
& $ - { \lambda_5 \over 12 m_\eta^4} (Y_u)_{pr} \left[Y_\eta^* m_N G_4(x)Y_\eta^\dagger  \right]_{st}$ \\
\hline
\end{tabular}
}
}
\caption{The matching result of dim-5 and dim-7 operators in the standard bases~\cite{Weinberg:1979sa,Liao:2020roy}. Seven out of twelve dim-7 LNV but  baryon-number-conserving operators are generated.}
\label{tab:radssaw_dim57basis}
\end{table}

Using the above EoMs together with IBP and Fierz relations, we can reduce the operators mentioned above to the standard basis. The final matching results for the LNC dim-6 and LNV dim-5 and dim-7 operators are tabulated in \mtab{tab:radssaw_Wbasis} and \mtab{tab:radssaw_dim57basis} respectively. We provide some details of reduction in appendix \ref{app:opered}. In the tables we introduced some loop functions appearing in \mtab{tab:radssaw_Gbasis} for brevity. 
Denoting the ratio of the two masses squared $x\equiv m_N^2/m_\eta^2$ where the flavor index of $N$ is not shown for brevity, we define the five loop functions associated with the dim-6 LNC operators, 
\begin{subequations}
\begin{eqnarray}
F_1(x) & \equiv & { 1 - 4 x + 3(1- 2\ln x )x^2  \over (1 -x )^3},
\\%
F_2(x) & \equiv &  {  1 - 6 x + 3(1- 2\ln x )x^2 +2 x^3 \over (1 -x )^4},
\\%
F_3(x) & \equiv &  {2 - 9 x + 18 x^2 - (11 - 6 \ln{x})x^3 \over 2 (1-x)^4},
\\%
F_{4,1}(x_v, x_w) & \equiv & - {\sqrt{ x_vx_w} \over 4}\left\{ \left[ { x_v \ln{x_v} \over  (x_v -x_w )(1 - x_v)^2} +v\leftrightarrow w\right]  + {  1 \over (1 - x_v) (1- x_w) } \right\},
  \\
F_{4,2}(x_v, x_w) & \equiv & {1\over 8}\left\{ \left[ { x_v^2 \ln{x_v}
 \over  (x_v -x_w )(1 - x_v)^2} +v\leftrightarrow w\right]  + {  1 \over (1 - x_v) (1- x_w) } \right\}.
\end{eqnarray}
\end{subequations}
Note that the functions $F_{4,1}(x_v, x_w)$ and $F_{4,2}(x_v, x_w)$ depend on two masses of $N_v$ and $N_w$. In the degenerate limit of $x_v=x_w=x$, they become $F_{4,i}(x, x)=F_{4,i}(x)$ where 
\begin{subequations}
\begin{eqnarray}
F_{4,1}(x)&\equiv &- {(2 +  \ln{x} )x - ( 2 - \ln{x}) x^2 \over 4 ( 1- x)^3},
\\
F_{4,2}(x)&\equiv &{1 + 2 x \ln{x} - x^2 \over 8(1 - x)^3}.
\end{eqnarray}
\end{subequations}
The four loop functions associated with the LNV dim-5 and dim-7 operators are, 
\begin{subequations}
\begin{eqnarray}
G_1(x) & \equiv & { 1 - (1 - \ln x) x  \over (1- x )^2} ,
\\%
G_2(x) &\equiv &{ 1 +  2 x  \ln x  - x^2 \over (1- x )^3},
\\%
G_3(x) & \equiv & { 1 +4(1 + \ln x) x-(5 - 2 \ln x )x^2  \over  (1- x )^4}  ,
\\%
G_4(x) & \equiv & { 1  -6 x  + 3 (1 - 2 \ln x ) x^2 + 2x^3  \over (1-x  )^4}.
\end{eqnarray}
\end{subequations}
Except for $F_{4,i}$, all functions are normalized to unity at $x=0$. 
We make a few comments concerning the matching result. First, the operators in blue in \mtab{tab:radssaw_Wbasis} and \mtab{tab:radssaw_dim57basis} are completely generated through the EoM operators in blue in \mtab{tab:radssaw_Gbasis}. Second, application of the EoMs also causes significant changes in the WCs for the operators already in the standard form, except for a few operators, i.e., $\calO_{W,HW,HB,HWB}$ in \mtab{tab:radssaw_Wbasis}, and $\calO_{LDH2}$ in \mtab{tab:radssaw_dim57basis}. And finally, application of EoMs for the two operators with a ${\color{cyan}\divideontimes}$ in \mtab{tab:radssaw_Gbasis} results in a further shift in the Higgs self-coupling $\hat\lambda_H$ in \meqn{eq:SMcorrection} with the final answer being 
\begin{eqnarray}
\hat \lambda_H \to \tilde \lambda_H
& \equiv & \hat \lambda_H + {1\over 16\pi^2} {1 \over 120} {\mu_H^2 \over m_\eta^2} [20(\lambda_4^2 + \lambda_5^2) - g_2^4]
\nonumber
\\
& = & \lambda_H
- {1\over 16\pi^2}\left\{ {L_\eta\over 2}( 2\lambda_3^2+2 \lambda_3\lambda_4 +\lambda_4^2 +\lambda_5^2 )
- {1 \over 120} {\mu_H^2 \over m_\eta^2} [20(\lambda_4^2 + \lambda_5^2) - g_2^4] \right\}.
\label{eq:newlambdaH}
\end{eqnarray}

\section{Phenomenology}
\label{sec:pheno}

From the SMEFT matching result in \mtab{tab:radssaw_Wbasis} and \mtab{tab:radssaw_dim57basis}, together with the matching results onto the LEFT given in~\cite{Jenkins:2017jig,Liao:2020zyx}, one can readily consider interesting physical processes to explore the parameter space of the model. 
Differently from all previous studies on the scotogeneric model~\cite{Kubo:2006yx, AristizabalSierra:2008cnr,Suematsu:2009ww,Adulpravitchai:2009gi,Toma:2013zsa,Vicente:2014wga,Huang:2018vcr,  Avila:2019hhv,deBoer:2021pon,Avila:2021mwg,Schmidt:2012yg,Ibarra:2016dlb,Hessler:2016kwm,Lindner:2016kqk,Baumholzer:2018sfb,Merle:2015ica,Kitabayashi:2018bye,Hugle:2018qbw,Borah:2018rca,Baumholzer:2019twf, Liu:2022byu,Borah:2020wut,Hundi:2022iva}, we here use the SMEFT approach to consider the heavy new physics implications for low energy physics, with both $m_\eta$ and $m_N$ being above the TeV scale. We will also make comparisons with the study in literature when necessary.

\subsection{Dim-5 and dim-7 LNV processes}

\begin{figure}
\centering
\includegraphics[width=8cm]{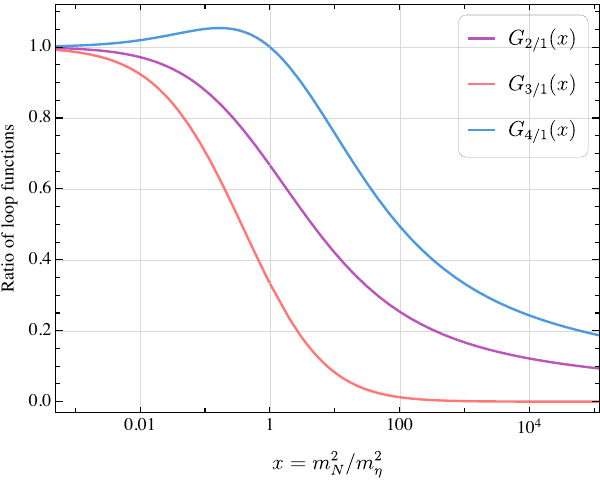}
\caption{Ratios of loop functions associated with LNV dim-5 and dim-7 matching coefficients as a function of the mass squared ratio $x$.}
\label{fig:d57LFratio}
\end{figure}

\noindent
{\bf\large Neutrino mass}:  The neutrino mass is generated from both the dim-5 Weinberg operator $\calO_{LH,5}$ and the dim-7 operator $\calO_{LH}$. By going to the Higgs phase $H \to v/\sqrt{2}$, we obtain the symmetric Majorana neutrino mass matrix as
\begin{eqnarray}
M^\nu_{pr} =
 {\lambda_5 v^2 \over 32\pi^2m_\eta^2 }
  \left\{Y_\eta^* m_N\left[G_1(x)- { (\lambda_3 + \lambda_4)v^2 \over 2 m_\eta^2}G_2(x)\right] Y_\eta^\dagger\right\}_{pr},
\end{eqnarray}
where the $G_1(x)$ part comes from the dim-5 Weinberg operator and the $G_2(x)$ part is due to the dim-7 contribution. The contributions associated with $G_4(x)$ loop function for dim-5 operator ${\cal O}_{LH,5}$ and dim-7 operator ${\cal O}_{LH}$ cancel out. These contributions come from the reduction of the last operator in \mtab{tab:radssaw_Gbasis} by the EoM of Higgs field in \meqn{eq:Heom}. After replacing the Higgs field by its vev $v/\sqrt{2}$, the pure scalar field sector in $D^2H$ relating to operators ${\cal O}_{LH,5}$ and ${\cal O}_{LH}$ vanishes due to $\mu_H^2= \lambda_H v^2$. 
We recall that a sum over $N_w$ is implied in the above, where the subscript enters $ m_{N_w}(Y^*_{\eta})_{pw}(Y^*_{\eta})_{rw}$ and $x_w=m_{N_w}^2/m_\eta^2$. In the limit of $\lambda_5\to 0$, our result for the dim-5 contribution differs from the original one given in Eq.\,(12) in \cite{Ma:2006km} by a factor of 2, 
but is consistent with the correct result reported in~\cite{Merle:2015ica}, as the latter was further confirmed by other independent calculations like \cite{Escribano:2020iqq}.
The dim-7 contribution is new and has not been considered before in the literature. To assess roughly its potential relevance, we consider the ratio of their contributions: 
\begin{eqnarray}
{ (\lambda_3 + \lambda_4)v^2   \over 2 m_\eta^2}{G_2(x) \over G_1(x)}
\sim 0.3 {\lambda_3 + \lambda_4 \over 10}\left( 1\,\si{TeV} \over m_\eta \right)^2 G_{2/1}(x),
\end{eqnarray}
where $G_{2/1}(x)\equiv G_2(x)/G_1(x)$ deceases monotonically from 1 at $x=0$ to $0.17$ at $x=10^3$ (\mfig{fig:d57LFratio}). Thus, for a relatively large $\lambda_{3,4}\sim\calO(1)-\calO(10)$ and  $m_\eta\sim\calO(\si{TeV})$, the dim-7 contribution could be non-negligible and should be included in a refined analysis. 

\noindent
{\bf\large Neutrino transition moment}: From the matching results in \mtab{tab:radssaw_dim57basis} one would expect a contribution to the Majorana neutrino electromagnetic moment from the operator $\calO_{LHW}^{pr}$. However,  this potential contribution vanishes since its WC $C_{LHW}^{pr}$ is symmetric in the flavor indices while the neutrino transition magnetic moment is antisymmetric. It is evident from Feynman diagrams as  there is no contributing diagram due to the $\mathbb{Z}_2$ symmetry and charge conservation.

\noindent
{\bf\large Effects of remaining dim-7 operators}: To see the relative size of LNV signals due to the matching results of other dim-7 operators, it is helpful to first compare the loop functions $G_{2,3,4}(x)$ in dim-7 WCs with $G_1(x)$ for the dim-5 Weinberg operator. In \mfig{fig:d57LFratio} we show the ratios of the three functions, $G_{i/1}(x)=G_i(x)/G_1(x)$ for $i=2,3,4$. As can be seen in the figure, $G_{2,3,4}(x)$ are almost always smaller than $G_1(x)$ in the whole range, with the only exception of $G_4(x)$ for $0\leq x\leq 1$ where the two  functions are comparable. Thus to explore their potential largest effects, we can approximately replace $G_{2,3,4}(x)$ by $G_1(x)$ and $\lambda_5 v^2/(32\pi^2m_\eta^2)[Y_\eta^*m_N G_1(x)Y_\eta^\dagger]_{pr}$ by the neutrino mass matrix $M^\nu_{pr}$. This results in the estimation: 
\begin{eqnarray}
&&|C_{LeHD}^{pr}| \lesssim {(M^\nu Y_e)_{pr}\over 2v^2 m_\eta^2};\quad
|C_{LHW}^{pr}|,\, |C_{LDH2}^{pr}| \lesssim  {M^\nu_{pr}\over 6v^2 m_\eta^2};
\nonumber
\\
&&|C_{\bar eLLLH}^{prst}|,\, |C_{\bar dQLLH1}^{prst}|,\, |C_{\bar QuLLH}^{prst}| \lesssim
{(Y_x)_{pr} M^\nu_{st}\over 6v^2 m_\eta^2},
\end{eqnarray}
where in the last line $Y_x =Y^\dagger_e,~Y^\dagger_d,~Y_u$ for the three different WCs in question. Clearly, these WCs are suppressed by the neutrino mass as well as the heavy scale $m_\eta$, and in some cases further suppressed by the SM Yukawa couplings. If we take $M_\nu\sim \calO(0.1\,\si{eV})$ and $m_\eta\sim\calO(1\,\si{TeV})$, the largest WC is
$|C_{\bar QuLLH}^{prst}| \lesssim 1/(10^4\,\si{TeV})^3(\si{TeV}/m_\eta)^2$, far below the current sensitivity of nuclear neutrinoless double-$\beta$ decay ($0\nu\beta\beta$)~\cite{Cirigliano:2017djv,Liao:2019tep}. We therefore conclude that the most interesting LNV processes like $0\nu\beta\beta$ are dominantly mediated by the neutrino mass term through the contribution of dim-5 and dim-7 Weinberg operators $\calO_{LH,5}$ and $\calO_{LH}$, and the effects from the remaining dim-7 operators can be safely neglected. This pattern of LNV signals being bound to the tiny neutrino mass is understandable from the fact that lepton number violation can only happen for  $\lambda_5Y_\eta^*Y_\eta^\dagger\ne 0$, to which the neutrino mass matrix is proportional. 

\subsection{LNC processes due to dim-6 operators}

\noindent
{\bf\large CDF $W$-mass anomaly}: The CDF Collaboration recently reported a measurement for the $W$ boson mass $m_W^{\tt CDF} = 80433.5\pm 9.4\,\si{MeV}$ \cite{CDF:2022hxs}, which is a 7$\sigma$ deviation from the SM prediction, $m_W^{\tt SM}= 80357 \pm4\,\si{MeV}$ \cite{Workman:2022ynf}. Combining the result with all previous measurements yields a new world average, $m_W^{\tt new\,ave} = 80417\pm 18\,\si{MeV}$ \cite{Workman:2022ynf}, which is still far away from the SM prediction. Assuming Gaussian error propagation, the CDF result implies the following relative size of the excess,
\begin{eqnarray}
&&{\delta m_W^2 \over m_W^2 }\Big|_{\tt CDF}\equiv {m_{W,\tt CDF}^{2} - m_{W,\tt SM}^{2} \over m_{W,\tt CDF}^{2} } =(0.95\pm 0.13) \times 10^{-3},
\nonumber
\\
&&{\delta m_W^2 \over m_W^2 }\Big|_{\tt new\,ave} =(0.75\pm 0.23) \times 10^{-3}.
\label{eq:mWex}
\end{eqnarray}
The anomaly has stimulated many studies, see for instance~\cite{Fan:2022dck,Bagnaschi:2022whn,deBlas:2022hdk,Strumia:2022qkt} and references therein. An investigation of the scotogenic model to explain the anomaly has been given in~\cite{Batra:2022pej}. We will work with the SMEFT approach by using the matching result obtained in the previous section to examine its implications. 

To study the correction to $m_W^2$ in SMEFT we proceed as in~\cite{Bagnaschi:2022whn} and employ the following most precisely known parameters as the input: 
\begin{eqnarray}
\alpha_{\rm em}^{-1}(m_Z) = 127.95, \quad
G_F = 1.16638 \times 10^{-5}\,\si{GeV}^{-2}, \quad
m_Z = 91.1876\,\si{GeV}.
\label{eq:EWinput}
\end{eqnarray}
Then the correction to $m_W^2$ enters through the modifications to the expressions of $m_Z$ and the Fermi constant $G_F$ measured in muon decay, which are mainly induced by the four dim-6 SMEFT  operators, 
\begin{eqnarray}
&&\calO_{HWB} = H^\dagger \sigma^I HW_{\mu\nu}^I B^{\mu\nu},\quad
\calO_{HD}  =|H^\dagger D_\mu H|^2,
\nonumber
\\
&&\calO_{ll}  =( \overline{L}\gamma_\mu L)( \overline{L}\gamma^\mu L),\quad
\calO_{H l}^{(3)}  =(H^\dagger i\overleftrightarrow{D_\mu^I} H)( \overline{L}\sigma^I \gamma^\mu L).
\end{eqnarray}
The first two affect the SM prediction on $m_Z$ through field diagonalization while the latter two lead to a correction to $G_F$. Defining the effective weak mixing angle by the input parameters as,
\begin{eqnarray}
\cos^2\theta_W\equiv {1\over 2} \left( 1 + \sqrt{1- {4\pi \alpha_{\rm em} \over \sqrt{2} G_F m_Z^2} } \right),
\end{eqnarray}
the correction to the pole mass $m_W$ relative to the SM prediction ($m_W^{\tt SM}=m_Z^2 \cos^2\theta_W$) takes the form, to the linear order in the WCs associated with the above operators, 
\begin{eqnarray}
{\delta m_W^2 \over m_W^2 }
& = &- {\sin2\theta_W\over \cos2\theta_W} {1 \over 8G_F}
 \left[ {\cos\theta_W \over \sin\theta_W}  C_{HD} +  4 C_{HWB} \right.
 \nonumber\\
&&\left. + {\sin\theta_W  \over \cos\theta_W}  \left(2C_{Hl}^{(3),ee} +2 C_{Hl}^{(3),\mu\mu} - C_{ll}^{\mu ee\mu} - C_{ll}^{e\mu \mu e} \right)
 \right].
\end{eqnarray}
If the $SU(3)^5$ global flavor symmetry is assumed, the above expression reduces to Eq.\,(2.3) given in~\cite{Bagnaschi:2022whn}.
Note that the usual electroweak oblique parameters $S$ and $T$~\cite{Peskin:1990zt} are related to $C_{HWB}$ and $C_{HD}$ via the relations $S = (4 \sin\theta_W\cos\theta_W/\alpha_{\rm em}) C_{WB}/(\sqrt{2}G_F)$ and 
$T = - (1/2 \alpha_{\rm em})C_{HD}/(\sqrt{2}G_F)$.
From the matching result in \mtab{tab:radssaw_Wbasis}, we obtain, 
\begin{eqnarray}
{\delta m_W^2 \over m_W^2 }  =
 {\tan2\theta_W \over 128\pi^2 G_F m_\eta^2}
 \left[ { (4\pi)^2 \alpha_{\rm em}^2 \over 15 \sin^3 2\theta_W}
+  {\cos\theta_W\over \sin\theta_W}
 {\lambda_4^2 - \lambda_5^2 \over 6}
-  {4\pi \alpha_{\rm em} \lambda_4 \over 3\sin2\theta_W}
+ 2 {\sin\theta_W\over \cos\theta_W}  \tilde C_{ll}^{e\mu \mu e}
\right],
\end{eqnarray}
where we have used the approximate relations $g_1 = e/\cos\theta_W$, $g_2 =e/\sin\theta_W$, and $\alpha_{\rm em}=e^2/(4\pi)$, as well as the abbreviation 
\begin{eqnarray}
 \tilde C_{ll}^{e\mu \mu e} &=& F_{4,1}(x_w,x_v) (Y_\eta)_{1w}  (Y_\eta^\T)_{w2} (Y_\eta^*)_{1v} (Y_\eta^\dagger)_{v2}
 \nonumber
 \\
&&- F_{4,2}(x_w,x_v)(Y_\eta)_{1w}(Y_\eta^\dagger)_{w1}(Y_\eta)_{2 v}  (Y_\eta^\dagger)_{v2},
\end{eqnarray}
where summation over $N_v,~N_w$ is implied. 

\begin{figure}
\centering
\includegraphics[width=8cm]{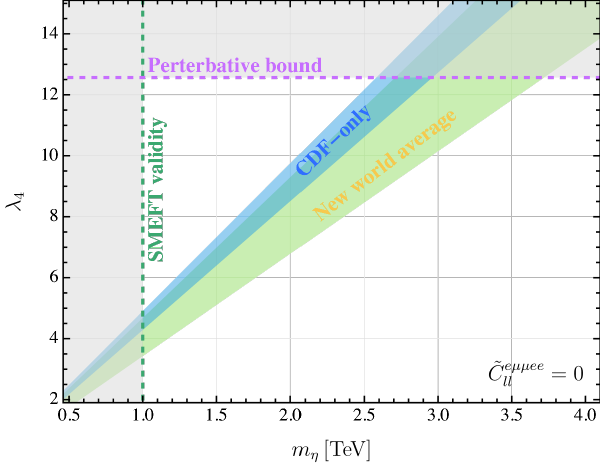}
\caption{The allowed region in the $m_\eta$-$\lambda_4$ plane for explaining the $W$ boson mass anomaly at the 1$\sigma$ level.}
\label{fig:mWa}
\end{figure}

Since $\lambda_5$ is related to the Majorana neutrino mass, it is naturally much smaller than unity to yield an $\calO(0.1\,\si{eV})$ neutrino mass for $\calO(\,\si{TeV})$ scale heavy masses. Dropping the $\lambda_5$ term and using the values in \meqn{eq:EWinput}, we obtain
\begin{eqnarray}
{\delta m_W^2 \over m_W^2 }
=
10^{-3} \left( \si{TeV} \over m_\eta \right)^2
\left(
1.6\times 10^{-4}
+0.046 \lambda_4^2
-0.006\lambda_4
+0.168 \tilde C_{ll}^{e\mu \mu e}
\right).
\end{eqnarray}
To reproduce the amount of excess in \meqn{eq:mWex} for $\calO(\si{TeV})$ scale $m_\eta$, it requires either a relatively large $\lambda_4$ or $ \tilde C_{ll}^{e\mu \mu e}$. Let us first consider the case with negligible $ \tilde C_{ll}^{e\mu \mu e}$. To explain the CDF anomaly, the allowed parameter space in the $m_\eta$-$\lambda_4$ plane is shown in \mfig{fig:mWa} for the two cases: the CDF-only and the new world average. 
It can be seen that $\lambda_4 \gtrsim 4$ is necessary to match the excess for $m_\eta \geq 1\,\si{TeV}$. This result basically agrees with the estimation given in \cite{Strumia:2022qkt}. On the other hand, the excess could be accommodated by $\tilde C_{ll}^{e\mu \mu e} \in [4.9,\,6.4]~([3.1,\,5.8])$ for the CDF-only (new world average) case as well. Since the loop functions are bound by $F_{4,1}\leq 1/24$ and $F_{4,2}\leq 1/8$, this would require $\calO({\rm few})$ 
Yukawa couplings to achieve a large enough $\tilde C_{ll}^{e\mu \mu e}$. 
 This may cause some tension with the constraints from the LFV processes ($\mu \to e \gamma$ and $\mu \to 3e$), and needs a careful analysis of the Yukawa sector.

\noindent
{\bf\large Lepton $\bm{g-2}$ and LFV decay $\bm{\ell_i \to \ell_j \gamma}$}:  These observables are connected to the dim-5 dipole operator in the LEFT,
\begin{eqnarray}
\calL_{e\gamma}
=C_{e\gamma}^{pr}\calO_{e\gamma}^{pr} +\hc
=  C_{e\gamma}^{pr}   \overline{e_{Lp} } \sigma_{\mu\nu}e_{Rr} F^{\mu\nu} +\hc,
\end{eqnarray}
where the WC $C_{e\gamma}^{pr}$ in LEFT are related to those in \mtab{tab:radssaw_Wbasis} by
\begin{eqnarray}
C_{e\gamma}^{pr} =
{v \over \sqrt{2}}\left( - \sin\theta_w C_{eW}^{pr} + \cos\theta_w C_{eB}^{pr}  \right)
= { e\, m_r \over 384\pi^2 } { 1 \over m_\eta^2}\left[ Y_\eta F_2(x)Y_\eta^\dagger \right]_{pr}.
\label{eq:WCdipole}
\end{eqnarray}
We have assumed that the SM Yukawa coupling matrix for charged leptons has been diagonalized so that  $M_e= v Y_e /\sqrt{2} = {\rm diag} (m_e, m_\mu,m_\tau)$. The decay width for the LFV process $\ell_i \to \ell_j \gamma$ is given by
\begin{eqnarray}
\Gamma_{\ell_i \to \ell_j \gamma}
= { m_{\ell_i}^3  \over 4\pi } \left( |C_{e\gamma}^{ji}|^2  + |C_{e\gamma}^{ij}|^2  \right)
= { \alpha_{\rm em} m_{\ell_i}^5 \over (384\pi^2)^2 m_\eta^4}
\left| \left[ Y_\eta F_2(x) Y_\eta^\dagger \right]_{ji}\right|^2,
\end{eqnarray}
where in the second step we have neglected the small correction proportional to ${ m_{\ell_j}^2/m_{\ell_i}^2 }$ from $|C_{e\gamma}^{ij}|^2$. The above decay width leads to the branching ratio 
\begin{eqnarray}
{ {\cal B}_{\ell_i \to \ell_j \gamma} \over {\cal B}_{\ell_i \to \ell_j \bar\nu_j \nu_i}^{\tt SM} }
&=&{\Gamma_{\ell_i \to \ell_j \gamma} \over \Gamma_{\ell_i \to \ell_j \bar\nu_j \nu_i}^{\tt SM} }
= { \alpha_{\rm em} \over 768\pi  G_F^2 m_\eta^4}
\left| \left[ Y_\eta F_2(x) Y_\eta^\dagger \right]_{ji}\right|^2
\nonumber
\\
&=& 2.2\times 10^{-8}\left( \si{TeV} \over m_\eta \right)^4
\left| \left[ Y_\eta F_2(x) Y_\eta^\dagger \right]_{ji}\right|^2,
\end{eqnarray}
where the SM prediction for the dominant decay width has been used, 
$\Gamma_{\ell_i \to \ell_j \bar\nu_j \nu_i}^{\tt SM} = G_F^2 m_{\ell_i}^5/(192\pi^3)$. 
In the heavy mass limit, our result agrees with those given in \cite{Kubo:2006yx,Vicente:2014wga}.
The current experimental upper bounds on these LFV processes are,
${\cal B}_{\mu \to e \gamma}^{\tt exp} \lesssim 4.2 \times 10^{-13}$ and ${\cal B}_{\tau \to e (\mu)\gamma}^{\tt exp} \lesssim 3.3(4.4) \times 10^{-8} $~\cite{Workman:2022ynf}, which implies
\begin{eqnarray}
{ {\cal B}_{\mu \to e \gamma} \over {\cal B}_{\mu\to e \bar \nu_e \nu_\tau}^{\tt SM} }
\lesssim 4.2 \times 10^{-13},  \quad
{ {\cal B}_{\tau \to e \gamma} \over  {\cal B}_{\tau \to e \bar \nu_e \nu_\tau}^{\tt SM} }
\lesssim 1.9\times 10^{-7},  \quad
{ {\cal B}_{\tau \to \mu \gamma} \over  {\cal B}_{\tau \to \mu \bar \nu_\mu \nu_\tau}^{\tt SM} }
\lesssim 2.5\times 10^{-7}.
\end{eqnarray}
As can be seen from the above, for $m_{\eta,N}\sim\calO(\si{TeV})$ and $\calO(1)$ Yukawa couplings,
the branching ratios of the LFV processes $\tau\to e(\mu)\gamma$ are below the current experimental bounds by an order of magnitude, whereas the $\mu \to e \gamma$ can probe the new physics scale up to tens of TeV for $\calO(1)$ Yukawa couplings.

While the electric dipole moment is not induced at one loop, there is a contribution to the anomalous magnetic moment of charged leptons: 
\begin{eqnarray}
\Delta a_\ell
= {m_\ell^2 \over 96\pi^2 m_\eta^2} \left[ Y_\eta F_2(x)Y_\eta^\dagger \right]_{\ell \ell}.
\end{eqnarray}
Considering $0<F_2(x) \leq 1$, the anomalous magnetic moment is approximately bound by
\begin{eqnarray}
|\Delta a_\ell|
\lesssim 1.2\times 10^{-11}
{m_\ell^2 \over m_\mu^2}
 \left({1\,{\rm TeV} \over m_\eta} \right)^2
\left|\left[ Y_\eta Y_\eta^\dagger \right]_{\ell \ell} \right|,
\end{eqnarray}
which cannot fill the gap between the SM prediction and the observed values in the Fermilab and BNL E821 experiments~\cite{Muong-2:2021ojo,Muong-2:2006rrc}, $\Delta a_\mu = a_\mu^{\tt exp} - a_\mu^{\tt SM} = 251(59)\times 10^{-11}$, for reasonably large Yukawa couplings. 

\noindent
{\large\bf{LFV decay $\mu \to3 e$}}: In the LEFT framework~\cite{Jenkins:2017jig,Crivellin:2017rmk}, the dominant contribution to $\mu \to 3e$ is from the long-distance dipole operator $\calO_{e\gamma}^{e\mu}$ through a virtual photon as well as the following dim-6 contact interactions,
\begin{subequations}
\begin{eqnarray}
\calO^{V,LL}_{e\mu ee} &=& (\overline{e_L}\gamma_\mu \mu_L)(\overline{e_L}\gamma^\mu e_L), \quad\,
\calO^{V,LR}_{e\mu ee} = (\overline{e_L}\gamma_\mu \mu_L)(\overline{e_R}\gamma^\mu e_R), \quad
\\
 \calO^{V,LR}_{eee\mu} &=& (\overline{e_R}\gamma_\mu \mu_R)(\overline{e_L}\gamma^\mu e_L),  \quad
\calO^{V,RR}_{e\mu ee} = (\overline{e_R}\gamma_\mu \mu_R)(\overline{e_R}\gamma^\mu e_R),  \quad
\\
\calO^{S,LL}_{e\mu ee} &=& (\overline{e_R}\mu_L)(\overline{e_R}e_L),  \quad\quad\quad\,
\calO^{S,RR}_{e\mu ee} = (\overline{e_L}\mu_R)(\overline{e_L}e_R).
\end{eqnarray}
\end{subequations}
Based on the matching result in \mtab{tab:radssaw_Wbasis} as well as the LEFT-SMEFT matching result in~\cite{Jenkins:2017jig}, 
we find only the first two dim-6 operators ($\calO^{V,LL}_{e\mu ee}, \calO^{V,LR}_{eee\mu}$) have non-negligible contributions, 
\begin{eqnarray}
C_{e\mu ee}^{V,LL} =  {1\over 16\pi^2 m_\eta^2}\left\{ 2\tilde C_{ll}^{e\mu ee} + {e^2\over 18}\left[Y_\eta F_3(x)Y_\eta^\dagger \right]_{e\mu} \right\}, \quad
C_{e\mu ee}^{V,LR} = {1\over 16\pi^2 m_\eta^2} {e^2\over 18}\left[Y_\eta F_3(x)Y_\eta^\dagger \right]_{e\mu}.
\end{eqnarray}
Rewriting the WC of the dipole operator in \meqn{eq:WCdipole} as 
$C_{e\gamma}^{e\mu} \equiv e\, m_\mu \tilde C_{e\gamma}^{e\mu}$, the decay width of $\mu\to 3e$ takes the form~\cite{Crivellin:2017rmk},
\begin{eqnarray}
{\Gamma_{\mu \to 3 e}} &=&
{m_\mu^5 \over 192\pi^3 }\bigg\{ e^4 |\tilde C_{e\gamma}^{e\mu}|^2
\left(8\ln{m_\mu \over m_e} - 11 \right)
+{1\over 8}\left(2 |C_{e\mu ee}^{V,LL} |^2 +|C_{e\mu ee}^{V,LR} |^2  \right)
\nonumber\\
&&
- e^2 \Re\left[\tilde C_{e\gamma}^{e\mu}(2 C_{e\mu ee}^{V,LL}+C_{e\mu ee}^{V,LR} )^* \right]
\bigg\},
\end{eqnarray}
where again we have neglected the small correction proportional to $m_e^2/m_\mu^2$ from $|\tilde C_{e\gamma}^{\mu e}|^2$.  In terms of branching ratios, we obtain
\begin{eqnarray}
{{\cal B}_{\mu \to 3 e} \over {\cal B}_{\mu \to e\bar \nu_e \nu_\mu}^{\tt SM}} 
&=&
{3e^4 \over 8 G_F^{2} }\bigg\{ {8 \over 3}|\tilde C_{e\gamma}^{e\mu}|^2
\left(8\ln{m_\mu \over m_e} - 11 \right)
+{1\over 3 e^4}\left(2 |C_{e\mu ee}^{V,LL} |^2 +|C_{e\mu ee}^{V,LR} |^2  \right)
\nonumber
\\
&&
- {8 \over 3e^2 }\Re\left[\tilde C_{e\gamma}^{e\mu}(2 C_{e\mu ee}^{V,LL}+C_{e\mu ee}^{V,LR} )^* \right]
\bigg\}.
\label{eq:mu23e}
\end{eqnarray}
Once again in the heavy mass limit, our result agrees with that in~\cite{Vicente:2014wga} upon 
the following correspondence of WCs between ours and theirs, 
\begin{eqnarray}
\tilde C_{e\gamma}^{e\mu} = {1\over 2} A_D, \quad
C_{e\mu ee}^{V,LL} = {1\over 2}e^2 (B + 2 A_{ND}),  \quad
C_{e\mu ee}^{V,LR} = e^2 A_{ND}.
\end{eqnarray}

Besides the processes discussed above, our matching results can also be applied to many other interesting processes, for instance, the LFV decays of the Higgs and $Z$ bosons, the LFV decays of heavy mesons and baryons, the $\mu$-$e$ conversion in nuclei, and non-standard neutrino  interactions, etc. This more complete phenomenological investigation deserves a future work. 

\section{Conclusion}
\label{sec:conc}

The scotogenic model is an economical extension of the SM that generates tiny neutrino mass at one-loop level, and at the same time naturally provides a dark matter candidate. The one-loop origin of neutrino mass implies a lower new physics scale than that of the three tree-level seesaws, which makes this model potentially testable at the future high energy colliders. Confronting with the current null experimental searches for weak scale new particles, it is appropriate to assume these new particles are well above the electroweak scale so that their indirect effects on low energy observables can be incorporated into an effective field theory where they have been integrated out. With this spirit, we employ the newly developed functional method to match the scotogenic model onto the standard model effective field theory up to dimension 7 for the case when both new particles $N$ and $\eta$ are heavy . Because of the $\mathbb{Z}_2$ symmetry, non-trivial matching only starts to appear at one loop. Our matching results are first organized in a Green basis (see \mtab{tab:radssaw_Gbasis}) in which the origin of the matched  operators can be relatively easily tracked. Then we use the SM equations of motion, the integration by parts relations, and various algebraic identities to recast the results in the standard bases for dim-5~\cite{Weinberg:1979sa}, dim-6~\cite{Grzadkowski:2010es}, and dim-7 operators~\cite{Liao:2020roy} (\mtab{tab:radssaw_Wbasis} and \mtab{tab:radssaw_dim57basis}). Finally, we apply our results to study implications on several interesting physical processes at low energy and make comparisons with those in the literature to further confirm our results. 

\section*{Acknowledgement}
\addcontentsline{toc}{section}{\numberline{}Acknowledgements}

This work was supported in part by the Grants No.~NSFC-12035008, No.~NSFC-11975130, No.~NSFC-12090064, by the Chung-Yao Chao Fellowship at Chinese Academy of Sciences Center for Excellence in Particle Physics (CCEPP), by the National Key Research and Development Program of China under Grant No. 2017YFA0402200, by the Guangdong Major Project of Basic and Applied Basic Research No. 2020B0301030008 and Science and Technology Program of Guangzhou No. 2019050001,  and by Shanghai Pujiang Program (20PJ1407800).

\newpage
\appendix

\section{Dim-5, dim-6, and dim-7 operator bases in the SMEFT}
\label{app:SMEFTbasis}

\begin{table}[!h]
\centering
\resizebox{\linewidth}{!}{
\renewcommand{\arraystretch}{1.3}
\begin{tabular}{|c|c|c|c|c|c|}
\hline
     \multicolumn{2}{|c|}{$\color{magenta}X^3$}
 &  \multicolumn{2}{c|}{ $\color{magenta}\psi^2H^3 +\hc$}
 &   \multicolumn{2}{c|}{ $\color{magenta} (\bar LL)(\bar LL)$} \\
 \hline%
     $\calO_G$ & $f^{ABC}G_{\mu}^{A\nu} G_\nu^{B\rho} G_\rho^{C\mu}$
     &  $\calO_{e H}$ & $(H^\dagger H)(\overline{L} eH)$
     & $\calO_{ll}$ & $(\overline{L}\gamma_\mu L)(\overline{L}\gamma^\mu L)$  \\
     $\calO_{\tilde G}$ & $f^{ABC}\tilde G_{\mu}^{A\nu} G_\nu^{B\rho} G_\rho^{C\mu}$
 &  $\calO_{u H}$ & $(H^\dagger H)(\overline{Q} u\tilde H)$
 &  $\calO_{qq}^{(1)}$ & $(\overline{Q}\gamma_\mu Q)(\overline{Q}\gamma^\mu Q)$ \\
 $\calO_{W}$ & $\epsilon^{IJK} W_{\mu}^{I\nu} W_\nu^{J\rho} W_\rho^{K\mu}$
 &  $\calO_{d H}$ & $(H^\dagger H)(\overline{Q} d H)$
 &  $\calO_{qq}^{(3)}$ & $(\overline{Q}\gamma_\mu \tau^I Q)(\overline{Q}\gamma^\mu \tau^I Q)$ \\
 \cline{3-4}
      $\calO_{\tilde W}$ & $\epsilon^{IJK} \tilde W_{\mu}^{I\nu} W_\nu^{J\rho} W_\rho^{K\mu}$
 &  \multicolumn{2}{c|}{$\color{magenta}\psi^2XH+\hc$}
 &  $\calO_{lq}^{(1)}$ & $(\overline{L}\gamma_\mu L)(\overline{Q}\gamma^\mu Q)$ \\
  \cline{1-4}
      \multicolumn{2}{|c|}{$\color{magenta}H^6$}
 &  $\calO_{eW}$ & $(\overline{L}\sigma^{\mu\nu}e) \tau^I H W^I_{\mu\nu}$
 &  $\calO_{lq}^{(3)}$ & $(\overline{L}\gamma_\mu \tau^I L)(\overline{Q}\gamma^\mu \tau^I Q)$ \\
\cline{1-2}\cline{5-6}
      $\calO_H$ & $(H^\dagger H)^3$
 &  $\calO_{eB}$ &  $(\overline{L}\sigma^{\mu\nu}e)H B_{\mu\nu}$
 &   \multicolumn{2}{c|}{$\color{magenta}(\bar RR)(\bar RR)$}   \\
 \cline{1-2} \cline{5-6}
     \multicolumn{2}{|c|}{$\color{magenta}H^4D^2$}
 &  $\calO_{uG}$ &  $(\overline{Q}\sigma^{\mu\nu}T^A u)\tilde H G^A_{\mu\nu}$
 &  $\calO_{ee}$ & $(\overline{e}\gamma_\mu e)(\overline{e}\gamma^\mu e)$ \\
  \cline{1-2}%
      $\calO_{H\Box}$ & $(H^\dagger H)\Box(H^\dagger H)$
 &  $\calO_{uW}$ & $(\overline{Q}\sigma^{\mu\nu} u)\tau^I \tilde H W^I_{\mu\nu}$
 &  $\calO_{uu}$ &$(\overline{u}\gamma_\mu u)(\overline{u}\gamma^\mu u)$ \\
   $\calO_{HD}$ & $(H^\dagger D_\mu H)^* (H^\dagger D^\mu H)$
 &  $\calO_{uB}$ & $(\overline{Q}\sigma^{\mu\nu} u)\tilde H B_{\mu\nu}$
 &  $\calO_{dd}$ & $(\overline{d}\gamma_\mu d)(\overline{d}\gamma^\mu d)$ \\
\cline{1-2}%
     \multicolumn{2}{|c|}{$\color{magenta}X^2 H^2$}
 &  $\calO_{dG}$ &  $(\overline{Q}\sigma^{\mu\nu}T^A d)H G^A_{\mu\nu}$
 &  $\calO_{eu}$ & $(\overline{e}\gamma_\mu e)(\overline{u}\gamma^\mu u)$ \\
\cline{1-2}%
     $\calO_{HG}$ & $H^\dagger H G^A_{\mu\nu}G^{A\mu\nu}$
 &  $\calO_{dW}$ &  $(\overline{Q}\sigma^{\mu\nu}d)\tau^I H W^I_{\mu\nu}$
 &  $\calO_{ed}$ & $(\overline{e}\gamma_\mu e)(\overline{d}\gamma^\mu d)$ \\
     $\calO_{H\tilde G}$ & $H^\dagger H \tilde G^A_{\mu\nu}G^{A\mu\nu}$
 &  $\calO_{dB}$ &  $(\overline{Q}\sigma^{\mu\nu}d) H B_{\mu\nu}$
 &  $\calO_{ud}^{(1)}$ & $(\overline{u}\gamma_\mu u)(\overline{d}\gamma^\mu d)$ \\
 \cline{3-4}
     $\calO_{H W}$ & $H^\dagger H  W^I_{\mu\nu}W^{I\mu\nu}$
 &   \multicolumn{2}{c|}{$\color{magenta}\psi^2 H^2D$}
 &  $\calO_{ud}^{(8)}$ & $(\overline{u}\gamma_\mu T^A u)(\overline{d}\gamma^\mu T^A d)$ \\
\cline{3-6}
      $\calO_{H\tilde W}$ & $H^\dagger H \tilde W^I_{\mu\nu}W^{I\mu\nu}$
 &  $\calO_{Hl}^{(1)}$  & $(H^\dagger i\overleftrightarrow{D_\mu} H)(\overline{L}\gamma^\mu L)$
 &   \multicolumn{2}{c|}{$\color{magenta}(\bar LL)(\bar RR)$}     \\
 \cline{5-6}%
      $\calO_{H B}$ & $H^\dagger H  B_{\mu\nu} B^{\mu\nu}$
 &  $\calO_{Hl}^{(3)}$  & $(H^\dagger i\overleftrightarrow{D_\mu^I} H)(\overline{L}\gamma^\mu \tau^I L)$
 &  $\calO_{le}$ & $(\overline{L}\gamma_\mu L)(\overline{e}\gamma^\mu e)$\\
      $\calO_{H \tilde B}$ & $H^\dagger H  \tilde B_{\mu\nu} B^{\mu\nu}$
 &  $\calO_{He}$  & $(H^\dagger i\overleftrightarrow{D_\mu} H)(\overline{e}\gamma^\mu e)$
 &  $\calO_{lu}$ & $(\overline{L}\gamma_\mu L)(\overline{u}\gamma^\mu u)$\\
      $\calO_{HWB}$ & $H^\dagger \tau^I H W^I_{\mu\nu} B^{\mu\nu}$
 &  $\calO_{Hq}^{(1)}$  & $(H^\dagger i\overleftrightarrow{D_\mu} H)(\overline{Q}\gamma^\mu Q)$
 &  $\calO_{ld}$ & $(\overline{L}\gamma_\mu L)(\overline{d}\gamma^\mu d)$\\
      $\calO_{H\tilde WB}$ & $H^\dagger \tau^I H \tilde W^I_{\mu\nu} B^{\mu\nu}$
 &  $\calO_{Hq}^{(3)}$  & $(H^\dagger i\overleftrightarrow{D_\mu^I} H)(\overline{Q}\gamma^\mu \tau^I Q)$
 &  $\calO_{qe}$ & $(\overline{Q}\gamma_\mu Q)(\overline{e}\gamma^\mu e)$\\
 &
 &  $\calO_{Hu}$  & $(H^\dagger i\overleftrightarrow{D_\mu} H)(\overline{u}\gamma^\mu u)$
 &  $\calO_{qu}^{(1)}$ & $(\overline{Q}\gamma_\mu Q)(\overline{u}\gamma^\mu u)$\\
 &
 &  $\calO_{Hd}$  & $(H^\dagger i\overleftrightarrow{D_\mu} H)(\overline{d}\gamma^\mu d)$
 &  $\calO_{qu}^{(8)}$ & $(\overline{Q}\gamma_\mu T^A Q)(\overline{u}\gamma^\mu T^A u)$\\
 &
 &  $\calO_{Hud}+\hc$  & $(\tilde H^\dagger i D_\mu H)(\overline{u}\gamma^\mu d)$
 &  $\calO_{qd}^{(1)}$ & $(\overline{Q}\gamma_\mu Q)(\overline{d}\gamma^\mu d)$\\
 &
 &  &
 &  $\calO_{qd}^{(8)}$ & $(\overline{Q}\gamma_\mu T^A Q)(\overline{d}\gamma^\mu T^A d)$ \\
 \cline{5-6}%
 &
 &  &
 &   \multicolumn{2}{c|}{$\color{magenta}(\bar LR)(\bar RL) +\hc$} \\
  \cline{5-6}%
 &
 &  &
 & $\calO_{ledq}$ & $(\overline{L}e)(\overline{d}Q)$ \\
\cline{5-6}%
 &
 &  &
 &   \multicolumn{2}{c|}{$\color{magenta}(\bar LR)(\bar LR) +\hc$} \\
  \cline{5-6}%
 &
 &  &
 & $\calO_{quqd}^{(1)}$ & $\epsilon_{ij}(\overline{Q^i} u)(\overline{Q^j}d)$ \\
  &
 &  &
 & $\calO_{quqd}^{(8)}$ & $\epsilon_{ij}(\overline{Q^i}T^A u)(\overline{Q^j}T^A d)$ \\
  &
 &  &
 & $\calO_{lequ}^{(1)}$ & $\epsilon_{ij}(\overline{L^i} e)(\overline{Q^j}u)$ \\
  &
 &  &
 & $\calO_{lequ}^{(3)}$ &\quad\quad  $ \epsilon_{ij}(\overline{L^i}\sigma_{\mu\nu} e)(\overline{Q^j}\sigma^{\mu\nu}u)$\quad\quad \\
\hline
\end{tabular}
}
\caption{The Warsaw basis of dim-6 operators that conserve baryon and lepton number in SMEFT \cite{Grzadkowski:2010es}. Note that in our convention the capital symbols $Q$ and $L$ are used to represent the quark and lepton doublets instead of $q$ and $l$.}
\label{tab:dim6opbasis}
\end{table}

At dim 5, there is a unique LNV operator related to the Majorana neutrino mass \cite{Weinberg:1979sa}
\begin{eqnarray}
\calO_{LH,5}^{pr}=\epsilon_{ij}\epsilon_{mn}(\overline{L^{\C,i}_p}L^m_r)H^jH^n,
\end{eqnarray}
plus its hermitian conjugate. \mtab{tab:dim6opbasis} and \mtab{tab:dim7opbasis} reproduce the bases of dim-6 and dim-7 operators  conserving the baryon number in SMEFT~\cite{Grzadkowski:2010es, Lehman:2014jma, Liao:2016hru}, respectively. The convention for fields is as follows: $L,~Q$ are the left-handed lepton and quark doublet fields, $u,~d,~e$ are the right-handed up-type quark, down-type quark, and charged lepton singlet fields, and $H$ denotes the Higgs doublet, respectively. For the dim-7 operators, they were first systematically studied in \cite{Lehman:2014jma}, and corrected by \cite{Liao:2016hru}. In \mtab{tab:dim7opbasis} we use the further improved basis in~\cite{Liao:2020roy} in which lepton flavor symmetry is more manifest and the operators containing both quark and lepton fields have a factorized quark-lepton current form. 

\begin{table}
\centering
\resizebox{\linewidth}{!}{
\renewcommand{\arraystretch}{1.1}
\begin{tabular}{|c|c|c|c|}
\hline
 \multicolumn{2}{|c|}{$\color{magenta}\psi^2H^4+\hc$} &  \multicolumn{2}{c|}{ $\color{magenta}\psi^2H^3D+\hc$}
\\
\hline
$\calO_{LH}$
& $\epsilon_{ij}\epsilon_{mn}(\overline{L^{\C,i}}L^m)H^jH^n(H^\dagger H)$
& $\calO_{LeHD}$
& $\epsilon_{ij}\epsilon_{mn}(\overline{L^{\C,i}}\gamma_\mu e)H^j(H^miD^\mu H^n)$
\\
\hline
\multicolumn{2}{|c|}{$\color{magenta}\psi^2H^2D^2+\hc$} &  \multicolumn{2}{c|}{$\color{magenta}\psi^2H^2X+\hc$}
\\
\hline
$\calO_{LDH1}$
& $\epsilon_{ij}\epsilon_{mn}(\overline{L^{\C,i}}\overleftrightarrow{D_\mu} L^j)(H^mD^\mu H^n)$
& $\calO_{LHB}$
& $ g_1\epsilon_{ij}\epsilon_{mn}(\overline{L^{\C,i}} \sigma_{\mu\nu}L^m)H^jH^nB^{\mu\nu}$
\\
$\calO_{LDH2}$
& $\epsilon_{im}\epsilon_{jn}(\overline{L^{\C,i}}L^j)(D_\mu H^m D^\mu  H^n)$
& $\calO_{LHW}$
& $g_2\epsilon_{ij}(\epsilon \tau^I)_{mn}(\overline{L^{\C,i}}\sigma_{\mu\nu}L^m)H^jH^nW^{I\mu\nu}$
\\
\hline
\multicolumn{2}{|c|}{$\color{magenta}\psi^4D+\hc$}  &   \multicolumn{2}{c|}{$\color{magenta}\psi^4H+\hc$}
\\
\hline
$\calO_{\overline{d}uLDL}$
& $\epsilon_{ij}(\overline{d}\gamma_\mu u)(\overline{L^{\C,i}}i\overleftrightarrow{D}^\mu L^j)$
& $\calO_{\overline{e}LLLH}$
& $\epsilon_{ij}\epsilon_{mn}(\overline{e}L^i)(\overline{L^{\C,j}}L^m)H^n$
\\
&  &
$\calO_{\overline{d}QLLH1}$
& $\epsilon_{ij}\epsilon_{mn}(\overline{d}Q^i)(\overline{L^{\C,j}}L^m)H^n$
\\
&  &
$\calO_{\overline{d}QLLH2}$
& $\epsilon_{ij}\epsilon_{mn}(\overline{d}\sigma_{\mu\nu}Q^i) (\overline{L^{\C,j}}\sigma^{\mu\nu} L^m)H^n$
\\
&  &
$\calO_{\overline{d}uLeH}$
& $\epsilon_{ij}(\overline{d}\gamma_\mu u)(\overline{L^{\C,i}}\gamma^\mu e)H^j$
\\
&  &
$\calO_{\overline{Q}uLLH}$
& $\epsilon_{ij}(\overline{Q}u)(\overline{L^{\C}}L^i)H^j$
\\
\hline
\end{tabular}
}
\caption{Basis of dim-7 lepton operators that conserve baryon number in SMEFT \cite{Liao:2020roy}. Here $D_\mu H^n$ is understood as $(D_\mu H)^n$, etc.}
\label{tab:dim7opbasis}
\end{table}

\section{Calculation of a power-type supertrace with double insertions of $\pmb{X}_{N\eta}$ and $\pmb{X}_{\eta N}$}
\label{sec:cal4fope}

The supertraces to calculate all have a general form $\str\left[f(P_\mu, \{U_k(x)\}) \right]$, where $P_\mu=iD_\mu$ with $D_\mu$ being the usual covariant derivative operator and $U_k$ is a set of $P_\mu$-independent functionals of classical fields and coupling constants. With the use of completeness relation and Baker-Campbell-Hausdorff formula, the CDE method leads to \cite{Cohen:2020qvb}
\begin{eqnarray}
{\rm Tr}\left[f(P_\mu, \{U_k(x)\}) \right]
=\int {\td^d q \over (2\pi)^d} \left\langle q \left| {\rm tr}\left[ f(P_\mu, \{U_k \}) \right] \right|q \right\rangle
= \int \td^d x
 \int {\td^d q \over (2\pi)^d}
{\rm tr}\left[ f(\bar P_\mu , \{\bar U_k \}) \right],
\end{eqnarray}
where the CDE transformed quantities are represented by a bar and take the following form,
\begin{eqnarray}
\bar P_\mu &=&\bar G_{\mu\nu} \tilde  \partial^{\nu}  - q_\mu, 
\nonumber
\\
\bar U_k &=&  \sum_{n=0}^\infty {1 \over n!} (P_{\mu_1}\dotsi P_{\mu_n} U_k)
 \tilde\partial^{\mu_1}\dotsi \tilde\partial^{\mu_n};
\nonumber
\\
i \bar G_{\mu\nu} &=&
 \sum_{n=0}{ n+1 \over (n+2)!} (P_{\mu_1}\dotsi P_{\mu_n} F_{\mu\nu})
\tilde\partial^{\mu_1}\dotsi \tilde\partial^{\mu_n},
\label{eq:CDEquant}
\end{eqnarray}
where $\tilde \partial^\mu \equiv \partial/\partial q_\mu$ stands for the partial derivative with respect to the loop momentum.

With the above formula, the supertrace with double insertions of $\pmb{X}_{N\eta}$ and $\pmb{X}_{\eta N}$ in \meqn{eq:allsuper} becomes,
\begin{eqnarray}
- {i \over 4}\str \left(\pmb{K}_N^{-1} \pmb{X}_{N\eta}^{[3/2]}\pmb{K}_\eta^{-1} \pmb{X}_{\eta N}^{[3/2]}\right)^2\Big|_{\tt hard}
= \int \td^4 x
{i \over 4}  \mu^{2\epsilon} \int{\td^d q \over (2\pi)^d}
{\rm tr}\left(\bar{\pmb{K}}_N^{-1} \bar{\pmb{X}}_{N\eta}^{[3/2]}\bar{\pmb{K}}_\eta^{-1} \bar{\pmb{X}}_{\eta N}^{[3/2]}\right)^2\Big|_{\tt hard},
\end{eqnarray}
where a minus sign is included due to the fermionic nature of the first propagator. Since the fermion $N$ is a gauge singlet, the CDE transformed inverse propagator $\bar{\pmb{K}}_N=-\slashed{q}+m_N$ with vanishing of $\bar G_{\mu\nu}^N$ due to $F_{\mu\nu}^N= 0$; while for the doublet $\eta$, $\bar{\pmb{K}}_\eta=(\bar G_{\mu\nu}^\eta\tilde \partial^\nu- q_\mu)^2- m_\eta^2$ with $F_{\mu\nu}^\eta=g_2 W_{\mu\nu}^I \sigma^I/2 + g_1 B_{\mu\nu} /2$ entering into the definition of $\bar G_{\mu\nu}^\eta$ in \meqn{eq:CDEquant}.
Denoting the obtained effective Lagrangian by $\calL_{4L}$ (with integration over spacetime $x$ implied on both sides), we have
\begin{eqnarray}
\calL_{4L} \overset{\tt CDE}{=}
 {i\over 4} \mu^{2\epsilon}\int{\td^d q \over (2\pi)^d}
{\rm tr}\left[ {1 \over q^2- m_N^2}( -\slashed{q} +m_N) \bar{\pmb{X}}_{N\eta}^{[3/2]}
{1 \over (\bar G_{\mu\nu}^\eta\tilde \partial^\nu- q_\mu)^2- m_\eta^2} \bar{\pmb{X}}_{\eta N}^{[3/2]} \right]^2\Big|_{\tt hard}.
\end{eqnarray}
Since each $\pmb{X}_{\eta N}^{[3/2]}$ contributes a lepton field $L(L^\C)$ from \meqn{eq:newX}, the supertrace will generate operators containing four $L(L^\C)$s without any Higgs field. Thus it cannot contribute to the LNV dim-7 operators but only LNC dim-6 ones involving four leptons. Therefore, up to dim-7 matching, we can remove the bars associated with CDE transformation and neglect the  $\bar G_{\mu\nu}^\eta $ part in the denominator, which leads to
\begin{eqnarray}
\calL_{4L}^{\tt dim \leq 7} & = &
 {i\over 4} \mu^{2\epsilon}\int{d^d q \over (2\pi)^d}
{\rm tr}\left( { -\slashed{q} +m_N \over q^2- m_N^2} \pmb{X}_{N\eta}^{[3/2]}
{1 \over q^2- m_\eta^2} \pmb{X}_{\eta N}^{[3/2]} \right)^2 \Big|_{\tt hard}
\nonumber
\\
&= &{i\over 4} \mu^{2\epsilon}\int{\td^d q \over (2\pi)^d}
{{\rm tr}\left[(-\slashed{q} +m_{N_v}) \pmb{X}_{N_v\eta}
\pmb{X}_{\eta N_w} (-\slashed{q} +m_{N_w}) \pmb{X}_{N_w\eta}
\pmb{X}_{\eta N_v} \right] \over (q^2- m_\eta^2)^2(q^2- m_{N_v}^2)(q^2- m_{N_w}^2)},
\label{eq:Lag4L}
\end{eqnarray}
where in the second step we have dropped the subscript ``{\tt hard}'' since the integrand already lives in the hard momentum region, and some flavor labels $(v,w)$ associated with the fermion $N$ are added. By taking into account the $\pmb{X}_{N\eta}$ and $\pmb{X}_{\eta N}$  in \meqn{eq:newX}, the trace part in the above can be calculated step by step as follows,
\begin{eqnarray}
{\rm trace}&\equiv&{\rm tr}\left[(-\slashed{q} +m_{N_v}) \pmb{X}_{N_v\eta}
\pmb{X}_{\eta N_w} (-\slashed{q} +m_{N_w}) \pmb{X}_{N_w\eta}
\pmb{X}_{\eta N_v} \right]
\nonumber
\\%
&\overset{(\ref{eq:newX})}{=}&
{\rm tr}\left\{
(-\slashed{q} +m_{N_v})
[(Y_\eta^\dagger)_{vp}  P_L L_p^i \bar L_r^i P_R  (Y_\eta)_{rw} +
 (Y_\eta^\T)_{vp}  P_R L_p^{i\C} \overline{L_r^{i\C}} P_L  (Y_\eta^*)_{rw}  ]
 \right.
 \nonumber
\\
&&\times\left.
(-\slashed{q} +m_{N_w})
[(Y_\eta^\dagger)_{ws}  P_L L_s^j \bar L_t^j  P_R  (Y_\eta)_{tv}  +
 (Y_\eta^\T)_{ws}  P_R L_s^{j\C} \overline{L_t^{j\C}} P_L  (Y_\eta^*)_{tv}  ]
 \right\},
\end{eqnarray}
where $(i,j)$ stand for the $SU(2)_L$ doublet indices and $(p,r,s,t,v,w)$ for the fermion flavor indices. Using the chiral projection property $P_R P_L=0$ and dropping the linear $q$ terms which vanish upon loop integration, it splits into two terms, 
\begin{eqnarray}
{\rm trace}=m_{N_v}m_{N_w}T_1+q_\alpha q_\beta T_2^{\alpha\beta}, 
\end{eqnarray}
where 
\begin{subequations}
\begin{eqnarray}
T_1&=& { \rm tr}\Big[(Y_\eta^\dagger)_{vp}  L_p^i \bar L_r^i (Y_\eta)_{rw}
 (Y_\eta^\T)_{ws}  L_s^{j\C} \overline{L_t^{j\C}} (Y_\eta^*)_{tv} 
\nonumber
\\
&&~~
+ (Y_\eta^\T)_{vp}  L_p^{i\C} \overline{L_r^{i\C}} (Y_\eta^*)_{rw}
 (Y_\eta^\dagger)_{ws} L_s^j \bar L_t^j  (Y_\eta)_{tv}  \Big]
 \nonumber
 \\
 & = &
- 2 (Y_\eta^\dagger)_{vp}  (Y_\eta)_{rw}  (Y_\eta^\T)_{ws} (Y_\eta^*)_{tv}
 \left[( \bar L_r^i  L_s^{j\C})(\overline{L_t^{j\C}} L_p^i)\right],
\\%
T_2^{\alpha\beta}&=&
{ \rm tr}\Big[ \gamma^\beta(Y_\eta^\dagger)_{vp}  L_p^i \bar L_r^i  (Y_\eta)_{rw}\gamma^\alpha
(Y_\eta^\dagger)_{ws}  L_s^j \bar L_t^j (Y_\eta)_{tv}   
\nonumber
\\
&&~~
+\gamma^\beta (Y_\eta^\T)_{vp}   L_p^{i\C} \overline{L_r^{i\C}} (Y_\eta^*)_{rw}\gamma^\alpha
 (Y_\eta^\T)_{ws}   L_s^{j\C} \overline{L_t^{j\C}}  (Y_\eta^*)_{tv} \Big]
\nonumber
\\
& = &
- 2(Y_\eta^\dagger)_{vp}  (Y_\eta)_{rw}(Y_\eta^\dagger)_{ws}(Y_\eta)_{tv}
 \left[( \bar L_r^i\gamma^\alpha L_s^j) (\bar L_t^j  \gamma^\beta L_p^i)  \right].
\end{eqnarray}
\end{subequations}
We have removed chiral projectors since our $L(L^\C)$ is left-handed, finished the trace in the spinor space, and reshuffled dummy indices to make the expressions more compact. Note that a minus sign has been included when interchanging a pair of fermion fields. Finally, we apply the FI in $T_1$, 
$2(\overline{\psi_{1L}} \psi^\C_{3L})(\overline{\psi^\C_{4L}}\psi_{2L})=( \overline{\psi_{1L}} \gamma^\mu \psi_{2L})( \overline{\psi_{3L}}  \gamma_\mu \psi_{4L})$, and make symmetric loop integration in $q_\alpha q_\beta T_2^{\alpha\beta}\to (q^2/d)g_{\alpha\beta}T_2^{\alpha\beta}$, to obtain 
\begin{subequations}
\begin{eqnarray}
T_1&=&-(Y_\eta)_{pw}  (Y_\eta^\T)_{ws} (Y_\eta^*)_{tv} (Y_\eta^\dagger)_{vr}
( \bar L_p\gamma_\mu L_r )( \bar L_s\gamma^\mu L_t ),
\\%
q_\alpha q_\beta T_2^{\alpha\beta} &\Rightarrow&
 - {2q^2 \over d} (Y_\eta)_{pw}(Y_\eta^\dagger)_{wt}(Y_\eta)_{sv}  (Y_\eta^\dagger)_{vr}
( \bar L_p\gamma_\mu L_r )( \bar L_s\gamma^\mu L_t ).
\end{eqnarray}
\end{subequations}
The loop integrals in \meqn{eq:Lag4L} can now be worked out to yield the loop functions $F_{4,1}(x,y)$ and $F_{4,2}(x,y)$. The end result is that the WC for the operator $( \bar L_p\gamma_\mu L_r )( \bar L_s\gamma^\mu L_t )$ is $\tilde C_{ll}^{prst}/(16\pi^2 m_\eta^2)$, where 
\begin{eqnarray}
 \tilde C_{ll}^{prst} & = &
F_{4,1}(x_v,x_w) (Y_\eta)_{pw}  (Y_\eta^\T)_{ws} (Y_\eta^*)_{rv} (Y_\eta^\dagger)_{vt}
\nonumber
\\
&& - F_{4,2}(x_v,x_w)  (Y_\eta)_{pw}(Y_\eta^\dagger)_{wt}(Y_\eta)_{sv}  (Y_\eta^\dagger)_{vr}, 
\end{eqnarray}
which is given in the last row of the dim-6 sector in \mtab{tab:radssaw_Gbasis}.

\section{Reduction into the standard basis}
\label{app:opered}

Using the integration by parts relations in \meqn{eq:scalaropered} and the EoMs in \meqn{eq:SMEoM}, we reduce the dim-6 non-standard basis operators (blue ones) in \mtab{tab:radssaw_Gbasis} to the Warsaw basis as follows,
\begin{subequations}
\begin{eqnarray}
 \partial_\mu B^{\mu\nu} \partial^\rho B_{\rho\nu} & \overset{\EoM}{\Rightarrow} &
 g_1^2
\left[
{1\over 6} \bar{Q} \gamma_\mu Q
+ {2\over 3} \bar{u} \gamma_\mu u
- {1\over 3} \bar{d} \gamma_\mu d
- {1\over 2} \bar{L} \gamma_\mu L
- \bar{e} \gamma_\mu e
+ {1\over 2}  H^\dagger i \overleftrightarrow{D_\mu} H
 \right]^2
 \nonumber
 \\
 & \Rightarrow &
{g_1^2 \over 4}\calO_{H\Box}
+ g_1^2 \calO_{H D}
- {g_1^2\over 2}\delta_{pr} \calO_{Hl}^{(1),pr}
-  g_1^2\delta_{pr}  \calO_{He}^{pr}
\nonumber
 \\
&&
+ {g_1^2\over 6} \delta_{pr}\calO_{Hq}^{(1),pr}
+ {2 \over 3} g_1^2 \delta_{pr}  \calO_{Hu}^{pr}
- {g_1^2 \over 3}\delta_{pr} \calO_{Hd}^{pr}
\nonumber
 \\
&&
+ {g_1^2 \over 4}\delta_{pr}\delta_{st}\calO_{ll}^{prst}
+ {g_1^2 \over 36 }\delta_{pr}\delta_{st} \calO_{qq}^{(1),prst}
- {g_1^2 \over 6}\delta_{pr}\delta_{st}\calO_{lq}^{(1),prst}
\nonumber
 \\
&&
+g_1^2 \delta_{pr}\delta_{st}\calO_{ee}
+ {4 \over 9}  g_1^2\delta_{pr}\delta_{st} \calO_{uu}^{prst}
+ { g_1^2 \over 9} \delta_{pr}\delta_{st} \calO_{dd}^{prst}
\nonumber
 \\
&&
- { 4 \over 3}  g_1^2\delta_{pr}\delta_{st} \calO_{eu}^{prst}
+ { 2\over 3} g_1^2 \delta_{pr}\delta_{st} \calO_{ed}^{prst}
- { 4\over 9} g_1^2\delta_{pr}\delta_{st}  \calO_{ud}^{(1),prst}
\nonumber
\\
& &
+ g_1^2 \delta_{pr}\delta_{st} \calO_{le}^{prst}
- { 2 \over 3} g_1^2\delta_{pr}\delta_{st}  \calO_{lu}^{prst}
+ { g_1^2 \over 3}\delta_{pr}\delta_{st} \calO_{ld}^{prst}
\nonumber
\\
& &
- { g_1^2 \over 3}\delta_{pr}\delta_{st} \calO_{qe}^{prst}
+ { 2 \over 9} g_1^2\delta_{pr}\delta_{st}  \calO_{qu}^{(1),prst}
- {  g_1^2 \over 9}\delta_{pr}\delta_{st}   \calO_{qd}^{(1),prst},
\label{eq:B2D2}
 \\%
D_\mu W^{I\mu\nu} D^\rho W^I_{\rho\nu} & \overset{\EoM}{\Rightarrow} &
{ g_2^2 \over 4}
\left[  \overline{Q}\sigma^I\gamma_\mu Q
+  \overline{L}\sigma^I\gamma_\mu L
 + H^\dagger  i \overleftrightarrow{D_\mu}^I H
 \right]^2
\nonumber
\\
& \Rightarrow &
\underline{\underline{ -  g_2^2 \mu_H^2 (H^\dagger H)^2 }}
 + 2 g_2^2 \lambda_H \calO_H
 + {3\over 4}g_2^2  \calO_{H\Box}
\nonumber
\\
&&
 +{g_2^2 \over 2}\delta_{pr} \calO_{Hl}^{(3),pr}
+ {g_2^2 \over 2}\delta_{pr} \calO_{Hq}^{(3),pr}
\nonumber
\\
& & + {g_2^2 \over 2} \big[ (Y_e)_{pr} \calO_{eH}^{pr}
+ (Y_u)_{pr} \calO_{uH}^{pr}
+ (Y_d)_{pr} \calO_{dH}^{pr}
+\hc \big]
\nonumber
\\
&& + {g_2^2 \over 4}( 2\delta_{pt}\delta_{rs} -\delta_{pr}\delta_{st}  )\calO_{ll}^{prst}
\nonumber
\\
&& + {g_2^2 \over 4}\delta_{pr}\delta_{st} \calO_{qq}^{(3),prst}
+ {g_2^2 \over 2}\delta_{pr}\delta_{st} \calO_{lq}^{(3),prst},
\\%
(H^\dagger H) (H^\dagger D^2 H) +\hc
&\overset{\EoM}{\Rightarrow} &
\underline{\underline{ 2\mu_H^2 (H^\dagger H)^2 }}
 -4\lambda_H \calO_H
\nonumber
\\
& &
- [ (Y_e)_{pr} \calO_{eH}^{pr}
+  (Y_u)_{pr} \calO_{uH}^{pr}
+  (Y_d)_{pr} \calO_{dH}^{pr}
+\hc ],
\\%
C_{pr} (H^\dagger H)(\overline{L_p} i \overleftrightarrow{ \slashed{D}}  L_r)
 & \overset{\EoM}{\Rightarrow} &
 (C Y_e)_{pr} \calO_{eH}^{pr} +\hc,
 \\%
C_{pr}  (H^\dagger\sigma^I H)(\overline{L_p} i \overleftrightarrow{ \slashed{D}}^I  L_r)
 & \overset{\EoM}{\Rightarrow } &
 (C Y_e)_{pr} \calO_{eH}^{pr} +\hc,
\\%
- C_{pr} \overline{L_p} i \overleftarrow{ \slashed{D}} i \slashed{D} i\slashed{D} L_r
& \overset{\EoM}{\Rightarrow}  &(Y_e^\dagger C Y_e)_{pr} (H^\dagger \overline{e_p})i\slashed{D}(e_r H)
\nonumber
\\
&=   &(Y_e^\dagger C Y_e)_{pr} \left[ (\overline{e_p}\gamma_\mu e_r)(H^\dagger i D^\mu H) +(H^\dagger H)(\overline{e_p}i\slashed{D}e_r)\right]
\nonumber
\\
&\overset{\IBP}{\Rightarrow} & {1\over 2} (Y_e^\dagger C Y_e)_{pr}  \left[ (\overline{e_p}\gamma_\mu e_r)
(H^\dagger i \overleftrightarrow{D^\mu} H) +(H^\dagger H) (\overline{e_p}i\overleftrightarrow{ \slashed{D}}e_r)\right]
\nonumber
\\
&\overset{\EoM}{\Rightarrow} &
{1\over 2} (Y_e^\dagger C Y_e)_{pr} \calO_{He}^{pr}
+{1\over 2} \left[  (Y_e Y_e^\dagger C Y_e)_{pr} \calO_{eH}^{pr} +\hc \right],
\\%
C_{pr} B^{\mu\nu}  \overline{L_p} \sigma_{\mu\nu} i  \slashed{D} L_r
& \overset{\EoM}{\Rightarrow} &
(C Y_e)_{pr} \calO_{eB}^{pr},
\\
C_{pr} D_\nu B^{\mu\nu}  \overline{L_p} \gamma_\mu L_r
& \overset{\EoM}{\Rightarrow} &
C_{pr}  g_1^2( \overline{L_p} \gamma_\mu L_r)
\bigg[
{1\over 6} \bar{Q} \gamma_\mu Q
+ {2\over 3} \bar{u} \gamma_\mu u
- {1\over 3} \bar{d} \gamma_\mu d
- {1\over 2} \bar{L} \gamma_\mu L
- \bar{e} \gamma_\mu e
 \nonumber
\\
&&+ {1\over 2}  H^\dagger i \overleftrightarrow{D_\mu} H
 \bigg]
 \nonumber
\\
& \Rightarrow &
  {g_1 \over 2}C_{pr} \calO_{Hl}^{(1), pr}
- {g_1\over 4} (C_{pr} \delta_{st} +C_{st} \delta_{pr} )\calO_{ll}^{prst}
+ {g_1\over 6} C_{pr} \delta_{st}\calO_{lq}^{(1),prst}
\nonumber
\\
& - &
 g_1C_{pr} \delta_{st}\calO_{le}^{prst}
+ { 2 g_1\over 3} C_{pr} \delta_{st}\calO_{lu}^{prst}
 -{ g_1\over 3} C_{pr} \delta_{st}\calO_{ld}^{prst},
\\%
C_{pr} W^{I\mu\nu}  \overline{L_p} \sigma^I \sigma_{\mu\nu} i  \slashed{D} L_r
& \overset{\EoM}{\Rightarrow} &
(C Y_e)_{pr} \calO_{eW}^{pr},
\\%
C_{pr} D_\nu W^{I\mu\nu}  \overline{L_p}\sigma^I  \gamma_\mu L_r
& \overset{\EoM}{=}  &
{g_2 \over 2} C_{pr}  (\overline{L_p}\sigma^I  \gamma^\mu L_r)
\left[  \overline{Q}\sigma^I\gamma_\mu Q
+  \overline{L}\sigma^I\gamma_\mu L
 + H^\dagger  i \overleftrightarrow{D_\mu}^I H
 \right]
 \nonumber
 \\
 & \Rightarrow &
 {g_2 \over 2} C_{pr} \calO_{Hl}^{(3), pr}
+ {g_2 \over 4} \big[2C_{pt} \delta_{rs}- C_{pr} \delta_{st} + (pr)\leftrightarrow (st)\big] \calO_{ll}^{prst}
 \nonumber
 \\
&& + {g_2 \over 2}  C_{pr} \delta_{st} \calO_{lq}^{(3),prst},\quad
\end{eqnarray}
\end{subequations}
where we have used a generic symbol $C_{pr}$ to stand for the WCs of the operators involving fermion fields. The WC of the operator $\calO_{ll}^{prst}$ is made symmetric using the flavor symmetry $\calO_{ll}^{prst}=\calO_{ll}^{stpr}$. The doubly underlined terms contribute to the dim-4 Higgs quartic interaction.

To reduce the two dim-7 EoM operators in  \mtab{tab:radssaw_Gbasis},  we need the identities,
\begin{eqnarray}
D^2 =  \slashed{D} \slashed{D} + {1\over2}\sigma^{\mu\nu}F_{\mu\nu},
\,
F_{\mu\nu} = i [D_\mu, D_\nu] =  g_1 Y B_{\mu\nu} +  g_2 T^I W^I_{\mu\nu},
\end{eqnarray}
where $Y$ is the hypercharge operator and $T^I$ the $SU(2)_L$ generators. For the lepton doublet field $L$ being acted on by $D^2$, $Y L =-{1\over 2}L$ and $T^I L = {1\over 2}\sigma^I L$, we have 
\begin{subequations}
\begin{eqnarray}
&&\epsilon_{ik} \epsilon_{jl}(\overline{L_p^{i,\C}} D^2 L_r^j) H^k  H^l
\nonumber
\\
&= &
  \epsilon_{ik}  \left[\epsilon_{jl} (\overline{L_p^{i,\C}} \slashed{D} \slashed{D}L_r^j)
- {g_1\over 4} \epsilon_{jl}(\overline{L_p^{i,\C}} \sigma^{\mu\nu}L_r^j)B_{\mu\nu}
- {g_2 \over 4}(\epsilon \sigma^I)_{jl}(\overline{L_p^{i,\C}} \sigma^{\mu\nu} L_r^j)W^I_{\mu\nu}
\right] H^k  H^l
\nonumber
\\
& \overset{\EoM}{\Rightarrow} & (Y_e)_{rs} \calO^{ps}_{LeHD}
- {1\over 4} \calO_{LHB}^{pr}
- {1\over 4}\calO_{LHW}^{pr},
\\%
&& \epsilon_{ik} \epsilon_{jl}(\overline{L_p^{i,\C}} L_r^j) H^k  D^2 H^l
\nonumber
\\
&  \overset{\EoM}{\Rightarrow} & \epsilon_{ik} \epsilon_{jl}( \overline{L_p^{i,\C}} L_r^j) H^k
\left[ \mu_H^2H^l - 2 \lambda_H (H^\dagger H) H^l + \epsilon_{lm} \overline{Q^m}  Y_u u
- \overline{d} Y^\dagger_d Q^l - \bar{e} Y^\dagger_e L^l \right]
\nonumber
\\
& = & \mu_H^2 \calO_{LH,5}^{pr} - 2\lambda_H\calO_{LH}^{pr}
- (Y_u)_{vw}  \calO_{\bar QuLLH}^{vwrp}
+ (Y^\dagger_d)_{vw}\calO_{\bar d QLLH1}^{vwrp}
+ (Y^\dagger_e)_{vw}\calO_{\bar e LLLH}^{vwrp},
\quad\quad
\end{eqnarray}
\end{subequations}
which lead to
\begin{eqnarray}
&&
S_{pr}\epsilon_{im} \epsilon_{jn}(\overline{L_p^{i,\C}} D^2 L_r^j)H^m  H^n
\nonumber
\\
&\overset{\EoM}{ \Rightarrow }&
(S Y_e)_{pr} \calO^{pr}_{LeHD}
- {1\over 4} S_{pr}\calO_{LHW}^{pr},
\\
&&S_{pr} \epsilon_{im} \epsilon_{jn}(\overline{L_p^{i,\C}} L_r^j) H^m D^2 H^n
\nonumber
\\
&\overset{\EoM}{ \Rightarrow}&
\underline{\underline{ \mu_H^2 S_{pr} \calO_{LH,5}^{pr} }}
- 2\lambda_H S_{pr}  \calO_{LH}^{pr}
\nonumber
\\
&+ & (Y^\dagger_e)_{pr} S_{st}\calO_{\bar e LLLH}^{prst}
+ (Y^\dagger_d)_{pr} S_{st} \calO_{\bar d QLLH1}^{prst}
 - (Y_u)_{pr} S_{st}  \calO_{\bar QuLLH}^{prst}.
\end{eqnarray}
Here $S_{pr}$ stands for the corresponding WC which is flavor symmetric, $S_{pr}=S_{rp}$, and the doubly underlined term contributes to the dim-5 Weinberg operator.

\renewcommand{\refname}{R\lowercase{eferences}}

\addcontentsline{toc}{section}{References}

\small


\end{document}